\documentclass{article}

\usepackage{amsmath}
\usepackage{amssymb}
\usepackage{graphicx}
\usepackage{pdflscape}

\def\be{\begin{eqnarray}}
\def\ee{\end{eqnarray}}
\def\nn{\nonumber}

\def\Tr{{\rm Tr}\,}

\def\l[{\phantom.[}

\numberwithin{equation}{section}

\hoffset= - 1.0in         

\title{{Evolution method and HOMFLY polynomials for virtual knots} \vspace{.2cm}}
\author{
{Ludmila Bishler}\thanks{{\small
\textit{Moscow State University}}; mila-bishler@mail.ru},
{Alexei Morozov}\thanks{{\small
\textit{ITEP, Moscow, Russia}} and \textit{National Research Nuclear University MEPhI}; morozov@itep.ru},
{Andrey Morozov}\thanks{{\small
\textit{Moscow State University} and \textit{ITEP, Moscow, Russia} and \textit{Laboratory of Quantum Topology, Chelyabinsk State University, Chelyabinsk, Russia} and \textit{National Research Nuclear University MEPhI}}; Andrey.Morozov@itep.ru}\,
and {Anton Morozov}\thanks{{\small \textit{
Moscow State University} and \textit{ITEP, Moscow, Russia}}; Anton.Morozov@itep.ru}
}
\date{\today}

\begin{document}

\maketitle

\vspace{-5cm}

\hfill ITEP/TH-35/14\\

\vspace{3.5cm}

\begin{abstract}
Following the suggestion of
\cite{MMMvirt}
to lift the knot polynomials for virtual knots and links
from Jones to HOMFLY,
we apply the evolution method to calculate them for an infinite
series of twist-like virtual knots and antiparallel 2-strand links.
Within this family one can check topological invariance
and understand how differential hierarchy is modified
in virtual case.
This opens a way towards a definition of colored
(not only cabled) knot polynomials,
though problems still persist beyond the first symmetric representation.
\end{abstract}



\tableofcontents


\section{Introduction}

Virtual knots were introduced in \cite{virtKauf}
as associated with the link diagrams, laying on non-trivial Riemann surfaces.
Since then the theory of virtual knots is slowly developing
\cite{virtfirst}.
From the point of view of knot polynomials
the biggest achievement until recently seems to be
evaluation of Jones polynomials for the simplest examples,
summarized in \cite{virtable}.
The main problem is the apparent breakdown of group-theoretical
approach of \cite{Wit,RT,MMMknots},
because the structure of representation theory is violated
by additional ``sterile'' vertex.

Recently in \cite{MMMvirt} it was suggested to overcome
this problem by applying the alternative calculus of \cite{DM3},
originally devised to substitute the overcomplicated construction
of Khovanov-Rozansky polynomials \cite{KR}.
Though this formalism is of the same (low) complexity level as
the Khovanov calculus for (super)Jones polynomials \cite{Kho},
for ordinary knots and links it immediately reproduces
HOMFLY polynomials for generic algebras $SL(N)$
-- but this is more or less a triviality
(non-trivial part of \cite{DM3} is a possibility to describe KR polynomials).
However, for virtual knots this formalism
(which supposedly remains absolutely the same as in the ordinary case)
is a real way out:  at this moment it is the only one,
providing HOMFLY polynomials for virtual knots and links.

Moreover, the very first attempt, made in \cite{MMMvirt},
demonstrated that the {\it answers} exhibit a structure,
which looks like just a minor
deformation of the
group-theoretical one -- and thus provides a hope for
building up a relevant modification of the approach of \cite{MMMknots}.

The goal of the present paper is to extend the first observations of \cite{MMMvirt}
to a little wider families of virtual knots and links.
This provides new evidence that the approach of \cite{DM3,MMMvirt}
indeed provides the topological invariant HOMFLY polynomials,
and also extends the observation about the needed group theory deformation.

We do not repeat here the details of the construction from \cite{DM3} and \cite{MMMvirt}:
it involves the standard conversion of link diagram into a hypercube \cite{Kho}
and subsequent reading of HOMFLY polynomial from data at the hypercube vertices.
Specifics of \cite{DM3} is a modification of one of the two resolutions,
which make vector spaces at hypercube vertices into factor-spaces,
and the split of hypercube construction into two steps:
that of the ``primary'' and then the "main" hypercube, which, in turn
is first built in the ``classical'' approximation and then ``quantized''.
Details can be found in \cite{DM3,MMMvirt} or will hopefully get clear from
consideration of the simple examples below.

\bigskip

The paper is organized as follows. In section \ref{2strator} we describe the {\it answers} for the
fundamental link polynomials
in the warming-up 2-strand torus case, partly known, partly new.
For their {\it derivations} see the next section \ref{2stratorussec}.

Then in sections \ref{twist} and \ref{twistcheck} we consider the family of twist knots,
ordinary and virtual -- the former are old results, the latter are new. The main new result here is eq.(\ref{Vbb}), it is a direct generalization of
the well-known (\ref{H-H}).
Like in the torus case we observe striking similarity between the answers
in ordinary and virtual cases -- despite the violation of representation theory
rules for the latter.
Moreover, the answers naturally exhibit the peculiar structure,
first observed in \cite{IMMMfe} for the figure-eight knot and
further promoted in \cite{evo} and \cite{artdiff} to the universal
``differential hierarchy'' structure, which probably reflect and materialize
the original observations in \cite{DGR} and became a powerful tool
for calculations of colored knot polynomials -- see also
\cite{diffappsfirst}.

In section \ref{diffvirt} we demonstrate that, like in the case of ordinary knots/links,
such universal differential structure (appropriately but universally modified)
persists for fundamental HOMFLY polynomials of {\it virtual knots}
beyond the family of twist knots.
Moreover, it can still be used for generalization to {\it colored}
polynomials -- which can be introduced by the standard cabling method.
This is, however, a tedious task, both technically and conceptually,
because now there is no natural projection on particular representation in the cable.
Using available data about {\it cabled} Jones polynomials for simple virtual knots,
we demonstrate in section \ref{colored} some {\it positive} evidence for the first
symmetric representation $[2]$ and some immediate {\it problems}
for the next symmetric representation $[3]$.
This is one of the obvious direction for further investigation.


\section{Evolution method for the two-stand braids \label{2strator}}

\subsection{Two parallel strands}

Following \cite{DM3}, we use
black and white vertices in oriented link diagrams to denote the two types of crossings:

\begin{picture}(100,60)(-70,-30)
\put(-39,-2){\mbox{${\cal R} =$}}
\put(-20,-20){\vector(1,1){40}}
\put(20,-20){\vector(-1,1){40}}
\put(0,0){\circle*{4}}
\put(29,-2){\mbox{$=$}}
\put(80,-20){\vector(-1,1){40}}
\put(40,-20){\line(1,1){17}}
\put(63,3){\vector(1,1){17}}
\put(148,-2){\mbox{${\cal R}^{-1} =$}}
\put(180,-20){\vector(1,1){40}}
\put(220,-20){\vector(-1,1){40}}
\put(200,0){\circle{4}}
\put(228,-2){\mbox{$=$}}
\put(240,-20){\vector(1,1){40}}
\put(280,-20){\line(-1,1){17}}
\put(257,3){\vector(-1,1){17}}
\end{picture}

\noindent
where the quantum ${\cal R}$-matrix and its inverse are inserted
in the RT formalism.
The big white circle is used to denote the "sterile" crossing,
where nothing is inserted:

\begin{picture}(100,60)(-100,-30)
\put(-39,-2){\mbox{${\cal P} =$}}
\put(-20,-20){\vector(1,1){40}}
\put(20,-20){\vector(-1,1){40}}
\put(0,0){\circle{7}}
%
%
\put(151,-2){\mbox{$\tilde{\cal P} =$}}
\put(180,-20){\vector(1,1){40}}
\put(180,20){\vector(1,-1){40}}
\put(200,0){\circle{7}}
\end{picture}


In \cite{MMMvirt} the fundamental HOMFLY polynomials were introduced for all the simplest
(triple-intersection) virtual knots from \cite{virtable}.
A more far-going example in \cite{MMMvirt} concerns the family of two-strand knots
of the shape

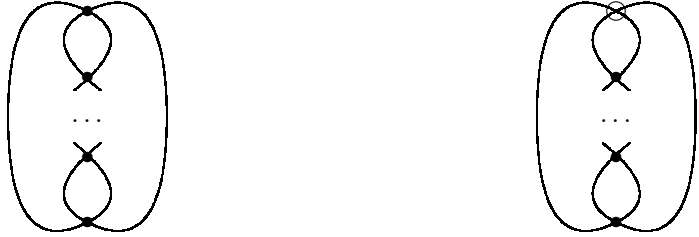
\begin{picture}(400,120)(-80,-60)
\qbezier(0,0)(0,55)(30,40)
\qbezier(0,0)(0,-55)(30,-40)
\qbezier(60,0)(60,55)(30,40)
\qbezier(60,0)(60,-55)(30,-40)
\qbezier(30,40)(50,30)(25,10)
\qbezier(30,40)(10,30)(35,10)
\qbezier(30,-40)(50,-30)(25,-10)
\qbezier(30,-40)(10,-30)(35,-10)
\put(24,-2){\mbox{$\ldots$}}
\put(30,40){\circle*{4}}
\put(30,-40){\circle*{4}}
\put(30,15){\circle*{4}}
\put(30,-15){\circle*{4}}
\qbezier(200,0)(200,55)(230,40)
\qbezier(200,0)(200,-55)(230,-40)
\qbezier(260,0)(260,55)(230,40)
\qbezier(260,0)(260,-55)(230,-40)
\qbezier(230,40)(250,30)(225,10)
\qbezier(230,40)(210,30)(235,10)
\qbezier(230,-40)(250,-30)(225,-10)
\qbezier(230,-40)(210,-30)(235,-10)
\put(224,-2){\mbox{$\ldots$}}
\put(230,40){\circle{7}}
\put(230,-40){\circle*{4}}
\put(230,15){\circle*{4}}
\put(230,-15){\circle*{4}}
\end{picture}

\noindent
In the left picture we have ordinary knots and links,
in the right -- the virtual ones appear.
It is sufficient to put just one sterile vertex,
because they commute with the ${\cal R}$-matrices and
two adjacent sterile vertices cancel each other.

It is assumed that there are $n$ black vertices in the left
picture and $n-1$ black vertices in the right one.
For negative $n$ the vertices are white rather than black.
When $n$ is odd, we have a knot, and it is independent of orientation.
When $n$ is even, we get a two-component link -- and it depends
on the mutual orientation of components: they can be either
parallel or antiparallel.
The two-strand knots (for odd $n$) are always parallel.

The main idea of the group-theoretic (RT) approach is that
${\cal R}$-matrices act as unit matrices in the irreducible
representations, only numerical eigenvalues should be taken into
account.
If only fundamental representations $[1]=\Box$ are ascribed to the
lines of the link diagram (reversion of arrow changes in to the
anti-fundamental $[1^{N-1}]=\overline{\Box}$), then there are just two
eigenvalues in the parallel channel and two in the anti-parallel:
\begin{equation}
\Box \otimes\Box \ = \ [2] + [11], \ \ \ \ \ \ \ \
{\cal R}_{[2]} = \frac{A}{q}\cdot{\rm Id}, \ \ \ \ \ \ \
{\cal R}_{[11]} = -{qA}\cdot{\rm Id}.
\label{parev}
\end{equation}
This implies that the HOMFLY polynomial in the parallel case is just
\begin{equation}
H^{[2,n]}_{_\Box} = \left(\frac{A}{q}\right)^{n}\cdot D_{[2]}\  +\
\Big(-{qA}\Big)^{n}\cdot D_{[11]} \ = \
\left(\frac{A}{q}\right)^{n}\cdot \frac{[N][N+1]}{[2]}\  +\
\Big(-{qA}\Big)^{n}\cdot \frac{[N][N-1]}{[2]},
\label{2braidknot}
\end{equation}
where $D_{[2]} = \frac{[N][N+1]}{[2]}$ and $D_{[11]} = \frac{[N][N-1]}{[2]}$
are quantum dimensions of the symmetric and antisymmetric representations
respectively. Here $A=q^N$ and the quantum number is defined as
\begin{equation}
[k] =\frac{\{q^k\}}{\{q\}} = \frac{q^k-q^{-k}}{q-q^{-1}},
\end{equation}
where $\{x\}=x-x^{-1}$.


Eq.(\ref{2braidknot}) can be obtained by the ``evolution method'' \cite{DMMSS,evo}:
as an ``average'' of the ${\cal R}$-matrix in ``representation form'',
\begin{equation}
H^{[2,n]}_{_\Box} = \
\left<\ \left(\begin{array}{cc} \frac{A}{q} & \\ & -{\footnotesize{qA}}\end{array}\right)^n\
\cdot {\cal P}_{{\rm 2-strands}}\right> \ = \ \Tr_{_{\Box\otimes\Box}}\ \check {\cal R}^n,
\end{equation}
corresponding to the splitting of the diagram into the ``evolution braid'' and ``vacuum projector'':

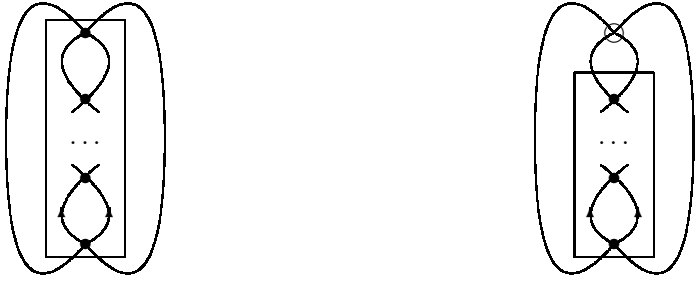
\begin{picture}(400,120)(-80,-60)
\qbezier(0,0)(0,75)(30,40)
\qbezier(0,0)(0,-75)(30,-40)
\qbezier(60,0)(60,75)(30,40)
\qbezier(60,0)(60,-75)(30,-40)
\qbezier(30,40)(50,30)(25,10)
\qbezier(30,40)(10,30)(35,10)
\qbezier(30,-40)(50,-30)(25,-10)
\qbezier(30,-40)(10,-30)(35,-10)
\put(24,-2){\mbox{$\ldots$}}
\put(30,40){\circle*{4}}
\put(30,-40){\circle*{4}}
\put(30,15){\circle*{4}}
\put(30,-15){\circle*{4}}
\put(15,-45){\line(1,0){30}}
\put(15,45){\line(1,0){30}}
\put(15,-45){\line(0,1){90}}
\put(45,-45){\line(0,1){90}}
\put(21,-28){\vector(0,1){2}}
\put(39,-28){\vector(0,1){2}}
\qbezier(200,0)(200,75)(230,40)
\qbezier(200,0)(200,-75)(230,-40)
\qbezier(260,0)(260,75)(230,40)
\qbezier(260,0)(260,-75)(230,-40)
\qbezier(230,40)(250,30)(225,10)
\qbezier(230,40)(210,30)(235,10)
\qbezier(230,-40)(250,-30)(225,-10)
\qbezier(230,-40)(210,-30)(235,-10)
\put(224,-2){\mbox{$\ldots$}}
\put(230,40){\circle{7}}
\put(230,-40){\circle*{4}}
\put(230,15){\circle*{4}}
\put(230,-15){\circle*{4}}
\put(215,-45){\line(1,0){30}}
\put(215,25){\line(1,0){30}}
\put(215,-45){\line(0,1){70}}
\put(245,-45){\line(0,1){70}}
\put(221,-28){\vector(0,1){2}}
\put(239,-28){\vector(0,1){2}}
\end{picture}

The idea of the evolution method is that whenever there is a box
of this type in a link diagram, with an $n$-point two-strand braid inside,
the dependence of HOMFLY polynomial on $n$ is fully described by the
linear combination
\begin{equation}
\alpha \cdot \left(\frac{A}{q}\right)^{n} \  +\
\beta\cdot \Big(-{qA}\Big)^{n},
\end{equation}
where $\alpha$ and $\beta$ depend on the "outside" of the box,
but does not depend on $n$.
These coefficients can be calculated if the answer is known for some two
values of $n$, e.g. $n=0$ and $n=\pm 1$.

\bigskip

In the particular case of the 2-strand parallel braid
\begin{equation}
H^{[2,0]}_{_\Box}=[N]^2\ \ \ \ \text{ and }\ \  \ H^{[2,\pm 1]}_{\Box}=[N]
\end{equation}
(two and one unknot respectively),
thus the coefficients $\alpha$ and $\beta$ are equal to $\frac{[N][N\pm 1]}{[2]}$,
as stated in   (\ref{2braidknot}).

\bigskip

Similarly, one can find the coefficients $\alpha$ and $\beta$
for the right diagram, i.e. for \text{virtual} knots an links:
\be
V^{[2,1]}_{_\Box}
\ = \ \Tr_{_{\Box\otimes\Box}}\ \check {\cal R}^n{\cal P} \ =
\left(\frac{A}{q}\right)^{n-1}\cdot \left(\frac{[N][N-1]}{[2]}+[N]\right)\  -\
\Big(-{qA}\Big)^{n-1}\cdot \frac{[N][N-1]}{[2]}
\label{2braidvirt}
\ee
in accordance with \cite{MMMvirt}.
We denote HOMFLY polynomials for virtual knots by $V$, to emphasize that these
are still new, under-investigated quantities with a more shaky status.

\bigskip

Operator ${\cal P}$ stands for the ``sterile'' vertex, and it breaks the
group-representation structure.
The most interesting fact is that
this breaking is not as radical as it could:
(\ref{2braidvirt}) differs from (\ref{2braidknot})
by the substitution
\begin{equation}
\begin{array}{lcl}
D_{[2]}=\cfrac{[N][N+1]}{[2]}  & \longrightarrow & \cfrac{[N][N-1]}{[2]}+[N],
\\[0.5cm]
D_{[11]}=\cfrac{[N][N-1]}{[2]}  & \longrightarrow & -\cfrac{[N][N-1]}{[2]}.
\end{array}
\end{equation}
These are the intriguing formulae which imply that HOMFLY polynomials
for virtual knots and links {\it can} one day acquire a representation-theory
interpretation.

To avoid possible confusion, we note that the middle part of eq.(\ref{2braidvirt})
is somewhat symbolic -- no generally-applicable matrix representation is known for
operators ${\cal R}$ and ${\cal P}$ together, thus one can not use this type of formulae
for practical calculations with virtual knots and links.
The only possibility at this moment is to use an
alternative approach of \cite{DM3} -- as suggested in \cite{MMMvirt}.
This is how the r.h.s. of (\ref{2braidvirt}) is actually derived.

\subsection{Two antiparallel strands \label{antipsec}}

In the antiparallel case the two product of two representations is
a combination of adjoint ${\rm Adj}= [21^{N-2}]$ and a singlet $[0]$:
\begin{equation}
\Box \otimes\overline{\Box} \ = \ {\rm Adj} + [0], \ \ \ \ \ \ \ \
{\cal R}_{{\rm Adj}} = A\cdot{\rm Id}, \ \ \ \ \ \ \
{\cal R}_{[0]} = 1\cdot{\rm Id}.
\label{antiparev}
\end{equation}
For the antiparallel link ($n=2k$ is obligatory even in this case)

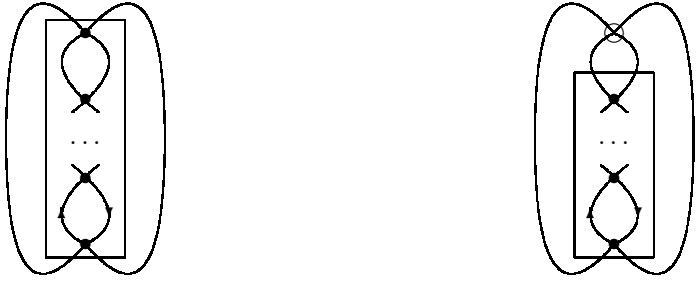
\begin{picture}(400,120)(-80,-60)
\qbezier(0,0)(0,75)(30,40)
\qbezier(0,0)(0,-75)(30,-40)
\qbezier(60,0)(60,75)(30,40)
\qbezier(60,0)(60,-75)(30,-40)
\qbezier(30,40)(50,30)(25,10)
\qbezier(30,40)(10,30)(35,10)
\qbezier(30,-40)(50,-30)(25,-10)
\qbezier(30,-40)(10,-30)(35,-10)
\put(24,-2){\mbox{$\ldots$}}
\put(30,40){\circle*{4}}
\put(30,-40){\circle*{4}}
\put(30,15){\circle*{4}}
\put(30,-15){\circle*{4}}
\put(15,-45){\line(1,0){30}}
\put(15,45){\line(1,0){30}}
\put(15,-45){\line(0,1){90}}
\put(45,-45){\line(0,1){90}}
\put(21,-28){\vector(0,1){2}}
\put(39,-28){\vector(0,-1){2}}
\qbezier(200,0)(200,75)(230,40)
\qbezier(200,0)(200,-75)(230,-40)
\qbezier(260,0)(260,75)(230,40)
\qbezier(260,0)(260,-75)(230,-40)
\qbezier(230,40)(250,30)(225,10)
\qbezier(230,40)(210,30)(235,10)
\qbezier(230,-40)(250,-30)(225,-10)
\qbezier(230,-40)(210,-30)(235,-10)
\put(224,-2){\mbox{$\ldots$}}
\put(230,40){\circle{7}}
\put(230,-40){\circle*{4}}
\put(230,15){\circle*{4}}
\put(230,-15){\circle*{4}}
\put(215,-45){\line(1,0){30}}
\put(215,25){\line(1,0){30}}
\put(215,-45){\line(0,1){70}}
\put(245,-45){\line(0,1){70}}
\put(221,-28){\vector(0,1){2}}
\put(239,-28){\vector(0,-1){2}}
\end{picture}

\noindent
we get
\begin{equation}
\gamma \cdot A^{2k} + \delta
\label{antipevo}
\end{equation}
with $n=2k$-independent $\gamma$ and $\delta$.

\bigskip

Note additional significant difference between evolutions
in parallel and antiparallel cases.
In variance with

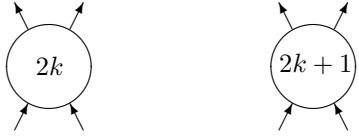
\begin{picture}(200,60)(-170,-27)
\put(0,0){\circle{30}}
\put(-8,14){\vector(-1,2){5}}
\put(8,14){\vector(1,2){5}}
\put(-13,-24){\vector(1,2){5}}
\put(13,-24){\vector(-1,2){5}}
\put(-5,-3){\mbox{$2k$}}
\put(100,0){\circle{30}}
\put(87,-24){\vector(1,2){5}}
\put(113,-24){\vector(-1,2){5}}
\put(92,14){\vector(-1,2){5}}
\put(108,14){\vector(1,2){5}}
\put(87,-2){\mbox{$2k+1$}}
\end{picture}

\noindent
for parallel strands,  the two diagrams

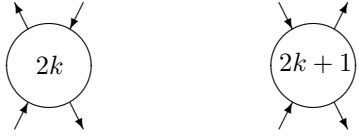
\begin{picture}(200,60)(-170,-30)
\put(0,0){\circle{30}}
\put(-8,14){\vector(-1,2){5}}
\put(8,-14){\vector(1,-2){5}}
\put(-13,-24){\vector(1,2){5}}
\put(13,24){\vector(-1,-2){5}}
\put(-5,-3){\mbox{$2k$}}
\put(100,0){\circle{30}}
\put(87,24){\vector(1,-2){5}}
\put(108,-14){\vector(1,-2){5}}
\put(87,-24){\vector(1,2){5}}
\put(108,14){\vector(1,2){5}}
\put(87,-2){\mbox{$2k+1$}}
\end{picture}

\noindent
with even and odd number of $R$-matrices in the 2-strand braid inside the circle
are essentially different. Therefore in the antiparallel case evolution works for insertion
of {\it even} number of vertices, while the two series with different parities
of vertex numbers can be quite different.
In the case of 2-strand antiparallel links this situation is just extreme:
the series with odd number of vertices simply does not exist.

Another word of caution is that the eigenvalues in (\ref{parev}) and (\ref{antiparev})
do not coincide for $N=2$, though from representation-theory point of view in this case
$\overline{[1]}\cong [1]$.
This is the feature of ``topological framing'', which
is adjusted to ensure topological invariance under the Reidemeister moves --
it requires that $A/q$ in (\ref{parev}) coincides with $1$ in (\ref{antiparev})
at $N=1$ rather than $(A/q,-Aq)$ with $(A,1)$  at $N=2$.

\bigskip

For ordinary links we get (see eq.(116) in \cite{evo}):
\begin{equation}
H^{[2,2k]}_{_{\Box\times\overline{\Box}}} \ =\  \Tr_{_{\Box\times\overline{\Box}}} \check {\cal R}^{2k}
\ =\  1 +A^{2k}\, [N-1]\,[N+1],
\label{antiplinks}
\end{equation}
while for the virtual links --
\begin{equation}
V^{[2,2k]}_{_{\Box\times\overline{\Box}}} \ =\  \Tr_{_{\Box\times\overline{\Box}}} \check {\cal R}^{2k}
\tilde{\cal P}
\ =\   1 + A^{2k-1}\, [N-1]\,\Big([N]+1\Big).
\label{antipvirt}
\end{equation}
This follows from evaluation of two particular examples
in the next section \ref{2stratorussec}.

Again, the difference between ordinary and virtual cases
-- caused by insertion of the sterile-vertex operator $\tilde{\cal P}$
(tilde means that it acts in another -- transposed -- channel) -- is reduced to the
intriguing change
\begin{equation}
\begin{array}{rcl}
D_{{\rm Adj}} = [N+1][N-1] & \longrightarrow & [N-1]\,\Big([N] + 1\Big)
\\[0.5cm]
D_{[0]} = 1 & \longrightarrow & 1
\end{array}
\end{equation}

\section{Torus two-strand links and knots \label{2stratorussec}}

In this section we derive the four formulae
(\ref{2braidknot}), (\ref{2braidvirt}), (\ref{antiplinks}) and (\ref{antipvirt})
for particular values of $n$ and $k$ (what is sufficient to determine the coefficients
$\alpha$, $\beta$, $\gamma$ and $\delta$ in the evolution formulae)
by the method of \cite{DM3,MMMvirt}.
In the case (\ref{2braidknot}) and (\ref{antiplinks}) of ordinary knots and links
this can be also done by the RT method \cite{MMMknots},
but for (\ref{2braidvirt}) and (\ref{antipvirt})
there are still no alternative derivations and even alternative {\it definition}
of what is the HOMFLY polynomial for virtual knots and links.

Later, in section \ref{twist} we consider another family -- of twist knots,
and there we encounter another, topologically equivalent,
realization of the same virtual knots.
HOMFLY polynomials will turn to be the same -- thus providing an evidence,
that our definition is indeed topologically invariant.

\subsection{Resolution and the Seifert cycles}

The main idea of Khovanov's approach is to consider all colorings
of the given link diagram at once and substitute the diagram by
a hypercube of its resolutions.

Namely, consider the diagram with all vertices black and resolve each vertex
in the following way:

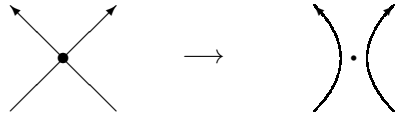
\begin{picture}(100,55)(-150,-25)
\put(-20,-20){\vector(1,1){40}}
\put(20,-20){\vector(-1,1){40}}
\put(0,0){\circle*{4}}
\put(45,-2){\mbox{$\longrightarrow$}}
\qbezier(95,-20)(115,0)(95,20)
\qbezier(125,-20)(105,0)(125,20)
\put(110,0){\circle*{2}}
\put(97,18){\vector(-1,1){2}}
\put(123,18){\vector(1,1){2}}
\end{picture}

Then for the ordinary link or knot the diagram decomposes into a collection
of non-intersecting (Seifert) cycles, while for virtual knots each sterile
vertex remains an intersection.
However, the resolutions of the parallel and antiparallel braids are
significantly different:

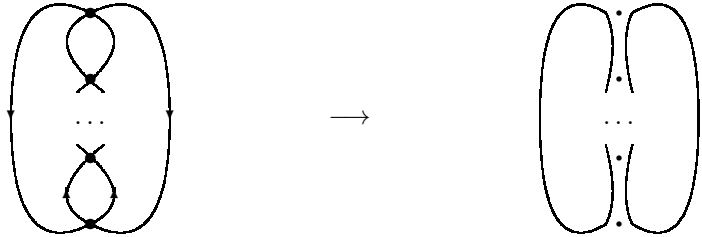
\begin{picture}(400,120)(-80,-60)
\qbezier(0,0)(0,55)(30,40)
\qbezier(0,0)(0,-55)(30,-40)
\qbezier(60,0)(60,55)(30,40)
\qbezier(60,0)(60,-55)(30,-40)
\qbezier(30,40)(50,30)(25,10)
\qbezier(30,40)(10,30)(35,10)
\qbezier(30,-40)(50,-30)(25,-10)
\qbezier(30,-40)(10,-30)(35,-10)
\put(24,-2){\mbox{$\ldots$}}
\put(30,40){\circle*{4}}
\put(30,-40){\circle*{4}}
\put(30,15){\circle*{4}}
\put(30,-15){\circle*{4}}
\put(0,1){\vector(0,-1){2}}
\put(60,1){\vector(0,-1){2}}
\put(21,-28){\vector(0,1){2}}
\put(39,-28){\vector(0,1){2}}
\put(120,-2){\mbox{$\longrightarrow$}}
\qbezier(200,0)(200,55)(225,40)
\qbezier(200,0)(200,-55)(225,-40)
\qbezier(260,0)(260,55)(235,40)
\qbezier(260,0)(260,-55)(235,-40)
\qbezier(225,40)(230,30)(225,10)
\qbezier(225,-40)(230,-30)(225,-10)
\qbezier(235,40)(230,30)(235,10)
\qbezier(235,-40)(230,-30)(235,-10)
%
%
\put(224,-2){\mbox{$\ldots$}}
\put(230,40){\circle*{2}}
\put(230,-40){\circle*{2}}
\put(230,15){\circle*{2}}
\put(230,-15){\circle*{2}}
\end{picture}

\noindent
i.e. in the parallel case we get just two Seifert cycles,
while in the antiparallel case there will be $n=2k$:

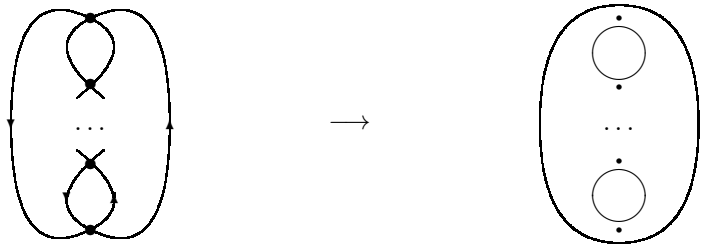
\begin{picture}(400,120)(-80,-60)
\qbezier(0,0)(0,55)(30,40)
\qbezier(0,0)(0,-55)(30,-40)
\qbezier(60,0)(60,55)(30,40)
\qbezier(60,0)(60,-55)(30,-40)
\qbezier(30,40)(50,30)(25,10)
\qbezier(30,40)(10,30)(35,10)
\qbezier(30,-40)(50,-30)(25,-10)
\qbezier(30,-40)(10,-30)(35,-10)
\put(24,-2){\mbox{$\ldots$}}
\put(30,40){\circle*{4}}
\put(30,-40){\circle*{4}}
\put(30,15){\circle*{4}}
\put(30,-15){\circle*{4}}
\put(0,0){\vector(0,-1){2}}
\put(60,0){\vector(0,1){2}}
\put(21,-28){\vector(0,-1){2}}
\put(39,-28){\vector(0,1){2}}
\put(120,-2){\mbox{$\longrightarrow$}}
\qbezier(200,0)(200,45)(230,45)
\qbezier(200,0)(200,-45)(230,-45)
\qbezier(260,0)(260,45)(230,45)
\qbezier(260,0)(260,-45)(230,-45)
\put(230,27){\circle{20}}
\put(230,-27){\circle{20}}
%
%
\put(224,-2){\mbox{$\ldots$}}
\put(230,40){\circle*{2}}
\put(230,-40){\circle*{2}}
\put(230,14){\circle*{2}}
\put(230,-14){\circle*{2}}
\end{picture}

\noindent
For virtual knots/links the number of Seifert cycles will be again different:
just one in the parallel case

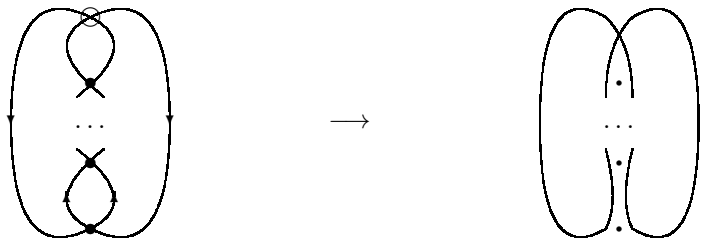
\begin{picture}(400,120)(-80,-60)
\qbezier(0,0)(0,55)(30,40)
\qbezier(0,0)(0,-55)(30,-40)
\qbezier(60,0)(60,55)(30,40)
\qbezier(60,0)(60,-55)(30,-40)
\qbezier(30,40)(50,30)(25,10)
\qbezier(30,40)(10,30)(35,10)
\qbezier(30,-40)(50,-30)(25,-10)
\qbezier(30,-40)(10,-30)(35,-10)
\put(24,-2){\mbox{$\ldots$}}
\put(30,40){\circle{7}}
\put(30,-40){\circle*{4}}
\put(30,15){\circle*{4}}
\put(30,-15){\circle*{4}}
\put(0,1){\vector(0,-1){2}}
\put(60,1){\vector(0,-1){2}}
\put(21,-28){\vector(0,1){2}}
\put(39,-28){\vector(0,1){2}}
\put(120,-2){\mbox{$\longrightarrow$}}
\qbezier(200,0)(200,55)(225,40)
\qbezier(200,0)(200,-55)(225,-40)
\qbezier(260,0)(260,55)(235,40)
\qbezier(260,0)(260,-55)(235,-40)
\qbezier(225,40)(235,30)(235,10)
\qbezier(225,-40)(230,-30)(225,-10)
\qbezier(235,40)(225,30)(225,10)
\qbezier(235,-40)(230,-30)(235,-10)
%
%
\put(224,-2){\mbox{$\ldots$}}
%
\put(230,-40){\circle*{2}}
\put(230,15){\circle*{2}}
\put(230,-15){\circle*{2}}
\end{picture}

\noindent
and $n-1=2k-1$ in the antiparallel case:

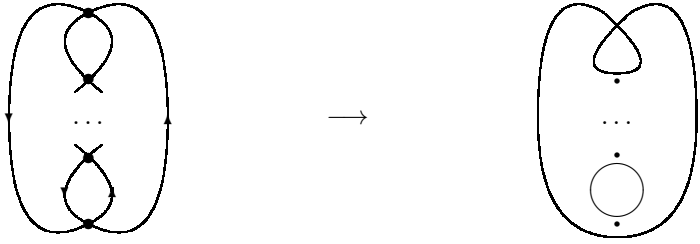
\begin{picture}(400,120)(-80,-60)
\qbezier(0,0)(0,55)(30,40)
\qbezier(0,0)(0,-55)(30,-40)
\qbezier(60,0)(60,55)(30,40)
\qbezier(60,0)(60,-55)(30,-40)
\qbezier(30,40)(50,30)(25,10)
\qbezier(30,40)(10,30)(35,10)
\qbezier(30,-40)(50,-30)(25,-10)
\qbezier(30,-40)(10,-30)(35,-10)
\put(24,-2){\mbox{$\ldots$}}
\put(30,40){\circle*{4}}
\put(30,-40){\circle*{4}}
\put(30,15){\circle*{4}}
\put(30,-15){\circle*{4}}
\put(0,0){\vector(0,-1){2}}
\put(60,0){\vector(0,1){2}}
\put(21,-28){\vector(0,-1){2}}
\put(39,-28){\vector(0,1){2}}
\put(120,-2){\mbox{$\longrightarrow$}}
\qbezier(200,0)(200,55)(225,40)
\qbezier(200,0)(200,-45)(230,-45)
\qbezier(260,0)(260,55)(235,40)
\qbezier(260,0)(260,-45)(230,-45)
\qbezier(225,40)(250,17)(230,17)
\qbezier(235,40)(210,17)(230,17)
\put(230,-27){\circle{20}}
%
%
\put(224,-2){\mbox{$\ldots$}}
%
\put(230,-40){\circle*{2}}
\put(230,14){\circle*{2}}
\put(230,-14){\circle*{2}}
\end{picture}

\subsection{Primary hypercube}

Above resolution of the link diagram stands at the
{\it main} vertex of the hypercube.
Other vertices correspond to alternative resolutions,
while edges of the hypercube are associated with the flip
of the resolution at a particular vertex of the link diagram.
Different versions of Khovanov formalism differ by the
choice of the second resolution.
The choice of \cite{DM3} -- generating knot polynomials
for arbitrary $N$ -- is rather complicated and is introduced
in three steps.
At the first step the flip to the second resolution is

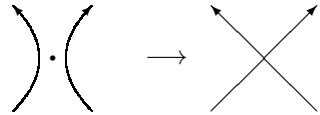
\begin{picture}(100,55)(-50,-25)
\qbezier(95,-20)(115,0)(95,20)
\qbezier(125,-20)(105,0)(125,20)
\put(110,0){\circle*{2}}
\put(97,18){\vector(-1,1){2}}
\put(123,18){\vector(1,1){2}}
\put(145,-2){\mbox{$\longrightarrow$}}
\put(170,-20){\vector(1,1){40}}
\put(210,-20){\vector(-1,1){40}}
\end{picture}

\noindent
i.e. introduces intersections of the Seifert cycles.

Hypercube introduced with the help of this flip is called {\it primary}
and in four above cases it looks rather different:

\bigskip

\centerline{
$
\begin{array}{ll}
\text{parallel ordinary case:}\ \ \ \
&
\begin{array}{|c|c|c|c|c|c|c|c|}
&&&&&& \underline{1} \ \ \text{for odd}\ n \\
\boxed{\underline{2}}&  n\times \underline{1} & C^2_n\times \underline{2} &
C^3_n \times \underline{1} & C^4_n \times \underline{2} & \ldots & \\
&&&&&& \underline{2}  \ \ \text{for even}\ n
\end{array}
\\ \nn \\
\text{antiparallel ordinary case:}
&
\begin{array}{|c|c|c|c|c|c|c|}
&&&&&&\\
\boxed{\underline{ n=2k}}&  n\times \underline{n-1} & C^2_n\times \underline{n-2} &
C^3_n \times \underline{n-3} &   \ldots & n\times \underline{1} &
\underline{2} \\
&&&&&&
\end{array}
\\ \nn \\
\text{parallel virtual case:}
&
\begin{array}{|c|c|c|c|c|c|c|c|}
&&&&&&\underline{1} \ \ \text{for odd}\ n \\
\boxed{\underline{1}}&  (n-1)\times \underline{2} & C^2_{n-1}\times \underline{1} &
C^3_{n-1} \times \underline{2} & C^4_{n-1} \times \underline{1} & \ldots& \\
&&&&&&\underline{2}  \ \ \text{for even}\ n
\end{array}
\\ \nn \\
\text{antiparallel virtual case:}
&
\begin{array}{|c|c|c|c|c|c|}
&&&&&\\
\boxed{\underline {n-1=2k-1}}&  (n-1)\times \underline{n-2} & C^2_{n-1}\times \underline{n-3} & \ldots
& (n-1)\times\underline{1} & \underline{2} \\
&&&&&
\end{array}
\end{array}
$
}

\bigskip

\noindent
Shown are the multiplicities of vertices at a given distance (number of flips)
from the main (Seifert) vertex with the given (underlined) number of cycles.
The main vertex is put into a box.
Binomial coefficients are the ordinary $C^i_n = \frac{n!}{i!(n-i)!}$.
The number of cycles in the {\it last} vertex is equal to the number of components
in the original link, i.e. in our case it is $\underline{1}$ for odd $n$
and $\underline{2}$ for even $n$.
In the virtual case the number of vertices in original l9ink diagram is $n-1$
rather than $n$, thus the dimension of hypercube is also less by one.

\subsection{The full classical hypercube}

The second step in the construction of \cite{DM3} is modification of the flip:

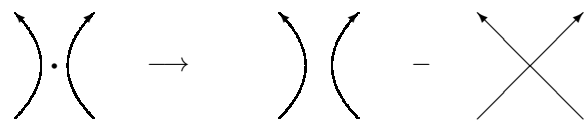
\begin{picture}(100,55)(-100,-25)
\qbezier(-5,-20)(15,0)(-5,20)
\qbezier(25,-20)(5,0)(25,20)
\put(10,0){\circle*{2}}
\put(-3,18){\vector(-1,1){2}}
\put(23,18){\vector(1,1){2}}
\put(45,-2){\mbox{$\longrightarrow$}}
\qbezier(95,-20)(115,0)(95,20)
\qbezier(125,-20)(105,0)(125,20)
\put(97,18){\vector(-1,1){2}}
\put(123,18){\vector(1,1){2}}
\put(145,-2){\mbox{$-$}}
\put(170,-20){\vector(1,1){40}}
\put(210,-20){\vector(-1,1){40}}
\end{picture}



\noindent
At $N=2$ the following procedure gives the same results
as Kauffman's orientation-violating flip \cite{KaufR}

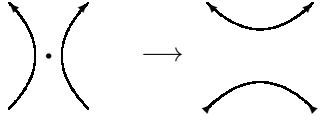
\begin{picture}(100,55)(-50,-25)
\qbezier(95,-20)(115,0)(95,20)
\qbezier(125,-20)(105,0)(125,20)
\put(110,0){\circle*{2}}
\put(97,18){\vector(-1,1){2}}
\put(123,18){\vector(1,1){2}}
\put(145,-2){\mbox{$\longrightarrow$}}
\put(170,-20){\vector(1,1){2}}
\put(210,-20){\vector(-1,1){2}}
\put(208,18){\vector(1,1){2}}
\put(172,18){\vector(-1,1){2}}
\qbezier(170,-20)(190,0)(210,-20)
\qbezier(170,20)(190,0)(210,20)
\end{picture}

\noindent
(there is no orientation-dependence at $N=2$, i.e. for Jones polynomials),
but for $N\neq 2$ there will be essential differences.

Since resolution is now a linear combination of two,
a hypercube vertex, which is at the distance $h$ from the main (Seifert) one
will be associated with a formal linear combination of $2^h$ resolved diagrams,
half of them with negative signs.
If the diagram has $\nu$ cycles, we substitute it by $N^\nu$ and take
a linear combination of these powers -- in Khovanov theory this can be
considered as a K-theory dimension (in the case of virtual knots/links
it can even be negative) of a ``factor-space'',
associated to the hypercube vertex.
Invariance with the widely familiar $N=2$ case, this association is
"non-local" -- depends on the $h$-dimensional {\it sub-hypercube}, connecting the given
vertex with the main (Seifert) vertex.

This construction provides what we call {\it classical} hypercube --
it can be easily restored once the {\it primary} one with marked (boxed)
Seifert vertex is given.
In our above examples we get:

\bigskip

{\footnotesize\centerline{
$
\begin{array}{ll}
\begin{array}{c}\text{parallel} \\ \text{ordinary case:}\end{array}
&
\begin{array}{|c|c|c|c|c|c|c|c|}
&&&&&&
\\
\boxed{\underline{N^2}}&  n\times \underline{N(N-1)} & C^2_n\times \underline{2N(N-1)} &
C^3_n \times \underline{2^2N(N-1)} & C^4_n \times \underline{2^3N(N-1)} & \ldots & 2^{n-1}N(N-1) \\
&&&&&&
\end{array}
\\ \nn \\
\begin{array}{c}\text{antiparallel} \\ \text{ordinary case:} \end{array}
&
\begin{array}{|c|c|c|c|c|c|c|}
&&&&&&\\
\boxed{\underline{N^n}}&  n\times \underline{N^{n-1}(N-1)} & C^2_n\times \underline{N^{n-2}(N-1)^2}
&C^3_n \times \underline{N^{n-3}(N-1)^3} &   \ldots & n\times \underline{N(N-1)^{n-1}} &
\underline{N(N-1)\,Y_{n-1}^{cl}} \\
&&&&&&
\end{array}
\\ \nn \\ \hline \nn \\
 \begin{array}{c} \text{parallel} \\ \text{virtual case:} \end{array}
&
\begin{array}{|c|c|c|c|c|c|c|}
&&&&&
\\
\boxed{\underline{N}}&  (n-1)\times \underline{-N(N-1)} & C^2_{n-1}\times \underline{-2N(N-1)} &
C^3_{n-1} \times \underline{-2^2N(N-1)}
& \ldots& -2^{n-1}N(N-1) \\
&&&&&
\end{array}
\\ \nn \\
 \begin{array}{c}\text{antiparallel} \\ \text{virtual case:}\end{array}
&
\begin{array}{|c|c|c|c|c|c|}
&&&&&\\
\boxed{\underline N^{n-1}}&  (n-1)\times \underline{N^{n-2}(N-1)} &
C^2_{n-1}\times \underline{N^{n-3}(N-1)^2} &
\ldots
& (n-1)\times\underline{N(N-1)^{n-2}} & \underline{N(N-1)\,y^{cl}_{n-2}} \\
&&&&&
\end{array}
\end{array}
$
}}

\bigskip

\noindent
Here
\begin{equation}
Y_{n-1}^{cl} = \frac{N^n - C^1_nN^{n-1}  + C^2_nN^{n-2}  - \ldots - C^{n-1}_n N
+ N^2}{N(N-1)} = \frac{(N-1)^n + N^2-1}{N(N-1)}
= \frac{(N-1)^{n-1}+N+1}{N}
\end{equation}
and
\begin{equation}
\begin{array}{r}
y_{n-2}^{cl} = \cfrac{N^{n-1} - C^1_{n-1}N^{n-2}  + C^2_{n-1}N^{n-3}  - \ldots - C^{n-2}_{n-1} N
- N^2}{N(N-1)} = \cfrac{(N-1)^{n-1} - N^2+1}{N(N-1)} =
\\[0.3cm]
= \cfrac{(N-1)^{n-2}-N-1}{N}.
\end{array}
\end{equation}


For even $n=2k$ the numerators at the r.h.s. are divisible by $N$.

Note that in the parallel virtual case most numbers are negative,
in the antiparallel virtual case this happens only to $y^{cl}_0 = -1$.

\subsection{The full quantum hypercube and HOMFLY polynomial}

The last step -- if we do not proceed further to Khovanov-Rozansky polynomials --
is to introduce $q$-dependence.
We call this procedure ``quantization'', in terms of vector spaces this means
conversion to $q$-graded spaces.

If quantum (graded) hypercube -- the set of numbers $\Big\{ {\cal N}_v\Big\}$ is known,
one can construct HOMFLY polynomials, by taking alternated sums
over the hypercube vertices, starting from {\it initial vertex} $v_c$,
defined by the coloring $c$ of original link diagram ${\cal L}_c$:
\begin{equation}
H^{{\cal L}_c}_{_\Box} = \frac{q^{n_\bullet(N-1)}}{(-q)^{n_\circ}}
\sum_{v\in {\cal H}({\cal L}} (-q)^{|h_v-h_c|} {\cal N}_v.
\end{equation}
Note that the only dependence on coloring is in the distance $|h_v-h_c|$
-- the hypercube ${\cal H}({\cal L})$ is the same for all colorings.

Usually, for a given knot/link its link diagram possesses alternative colorings,
associated with simpler links -- for which the answers are already known.
This can be used to specify the quantization procedure for the new,
yet unknown, items.

Most of quantities in above examples are quantized in a trivial way:
when the number is factorized into items $N-i$, each factor is just
substituted by a quantum number $[N-i]$.
In particular, this is sufficient for a full description of parallel families.

In antiparallel case one still needs quantization of $Y$ and $y$.
Applying the above-mentioned re-coloring trick, one can check that
naively-quantized are the differences:
\begin{equation}
Y_{n+1}^{cl}-Y_{n-1}^{cl} = \frac{(N-1)^{n+1}-(N-1)^{n-1}}{N} =
(N-1)^{n-1}(N-2) \ \longrightarrow \ [N-1]^{n-1}[N-2] = Y_{n+1} - Y_{n-1}
\end{equation}
and
\be
y_n^{cl}-y_{n-2}^{cl} = \frac{(N-1)^{n}-(N-1)^{n-2}}{N} =
(N-1)^{n-2}(N-2) \ \longrightarrow \ [N-1]^{n-2}[N-2] = y_n - y_{n-2}.
\ee
To this one should add initial conditions $Y_1 = [2]$ and $y_0=-1$.

\subsection{The four series \label{2stanswers}}

It remains to substitute all these prescriptions and deduce the answers
for our four examples.

\subsubsection{Ordinary parallel case}

For all black vertices we obtain

\begin{equation}\!\!
\begin{array}{l}
H^{[2,n]}_{_\Box} = q^{n(N-1)}\Big(
[N]^2 - qC^1_n[N][N-1] + q^2C^2_n [2][N][N-1] +
\ldots + (-q)^n [2]^{n-1}[N][N-1]\Big) =
\\[0.3cm]
= q^{n(N-1)}\left([N]^2-\frac{[N][N-1]}{[2]} + \frac{[N][N-1]}{[2]}\Big(1-q\,[2]\Big)^n\right) 
=\ \boxed{ q^{n(N-1)}\left( \frac{[N][N+1]}{[2]} +(-q^2)^n\cdot \frac{[N][N-1]}{[2]}\right). }
\end{array}
\label{2stratorus}
\end{equation}
This is exactly eq.(\ref{2braidknot}) from \cite{DM3}
-- a particular (2-strand) case of the celebrated
Rosso-Jones formula for torus knots and links \cite{RJ,DMMSS}.

\subsubsection{Ordinary antiparallel case}

For all black vertices we obtain (for $n=2k$)

\begin{equation}
\begin{array}{r}
H^{[2,2k]}_{_{\Box\times\overline{\Box}}} = q^{n(N-1)}\Big(
[N]^n - qC^1_n[N]^{n-1}[N-1] + q^2C^2_n[N]^{n-2}[N-1]^2 - \ldots -
\\[0.1cm] - q^{n-1}C^{n-1}_n[N][N-1]^{n-1} + q^n[N][N-1]Y_{n-1}
\Big) =
\\[0.3cm]
= q^{n(N-1)}\left\{\Big([N]-q\,[N-1]\Big)^n  + q^n[N][N-1]Y_{n-1} - q^n[N-1]^n\right\}=
\\[0.1cm]
= 1 + q^{nN)}[N-1]\Big([N]Y_{n-1} - [N-1]^{n-1}\Big),
\end{array}
\end{equation}
where we used the identity $[N]-q\,[N-1] = q^{1-N}$.

In order to understand what stands in the remaining bracket it
is sufficient to note that it is independent of $n$: the difference
$[N-1]^{n+1}-[N-1]^{n-1} = [N][N-1]^{n-1}[N-2]$
is exactly the same as $[N](Y_{n+1}-Y_{n-1})$.
Thus it remains to look at particular example of $n=2$,
when $[N]Y_{1} - [N-1] = [2][N]-[N-1]=[N+1]$.
Thus
\be
\boxed{
H^{[2,2k]}_{_{\Box\times\overline{\Box}}} = 1 + q^{2kN}[N+1][N-1],
}
\label{antiparrlink}
\ee
what is exactly (\ref{antiplinks}) from \cite{evo}.
Note that the only coincidence between (\ref{2stratorus}) and (\ref{antiparrlink})
is at $n=2$, the Hopf link is independent of orientation:
\be
H^{[2,2]}_{_{\Box\times\Box}} = \frac{q^{2N-2}[N]}{[2]}\Big([N+1]+q^4[N-1]\Big)
= 1 + q^{2N}[N-1][N+1] = H^{[2,2]}_{_{\Box\times\overline{\Box}}}.
\ee

In this example one needs to justify the quantization rule for $Y_{n-1}$,
thus it also deserves considering some other colorings.
If one vertex is white, then we should get the link with $n\longrightarrow n-2$.
On the other hand, this corresponds to taking as initial an vertex in the hypercube,
which is adjacent to the Seifert vertex.
Such checks were performed, at least partly, in \cite{DM3}.

\subsubsection{Virtual parallel case}

For all black vertices
\begin{equation}\!\!\!\!
\begin{array}{l}
V^{[2,n]}_{_\Box} = q^{(n-1)(N-1)}\Big([N] + qC^1_{n-1}[N][N-1] -q^2C^2_{n-1}[2][N][N-1]
+ q^3C^3_{n-1}[2]^2[N][N-1] - \ldots \Big)= \\[0.3cm]
=q^{(n-1)(N-1)} \left\{[N]+\cfrac{[N][N-1]}{[2]} - \frac{1}{[2]}(1-q\,[2])^{n-1}[N][N-1]\right\}
= \\[0.5cm]
\hfill\boxed{= q^{(N-1)(n-1)}\cdot \left( \frac{[N][N-1]}{[2]}+[N]\right)
- (-q^{N+1})^{n-1} \cdot\frac{[N][N-1]}{[2]}, }
\end{array}
\label{virt2stratorus}
\end{equation}
what is exactly the answer (\ref{2braidvirt}) from \cite{MMMvirt}.
Note the change of signs in most terms in the original sum, as compared to the case
of ordinary (non-virtual) knots and links -- it is caused by negativity of most
"dimensions"at hypercube vertices.
Also note that the combination $\frac{[N][N-1]}{[2]}+[N]$ can contain odd,
not only even powers of $q$ -- while for ordinary knots and links the knot
polynomials always depend on even powers of $q$.

\subsubsection{Virtual antiparallel case}

This time we get for the link diagram with all black vertices
(we remind that $n-1=2k-1$ is obligatory odd in this case):

\begin{equation}
\begin{array}{r}
V^{[2,2k]}_{_{\Box\times\overline{\Box}}} = q^{(n-1)(N-1)}\Big([N]^{n-1}
-qC^1_{n-1}[N]^{n-2}[N-1]+q^2C^2_{n-1}[N]^{n-3}[N-1]^2 - \ldots
\\[0.1cm]
\ldots
+ q^{n-2}C^{n-2}_{n-1}[N][N-1]^{n-2} - q^{n-1}[N][N-1]y_{n-2}\Big) =
\\[0.3cm]
%
= q^{(n-1)(N-1)}\left\{\Big([N]-q[N-1]\Big)^{n-1}
+q^{n-1}\Big([N-1]^{m-1}-[N][N-1]y_{n-2}\Big)
\right\}  =
\\[0.1cm]
=1 + q^{N(n-1)}[N-1]\Big([N-1]^{n-2} - [N]y_{n-2}\Big).
\end{array}
\end{equation}
%
%
Like in the ordinary antiparallel case the last bracket is independent of $n$
and for $n=2$ it is equal to $1+[N]$,
so that finally
\be
\boxed{
V^{[2,2k]}_{_{\Box\times\overline{\Box}}} =
1 + q^{(2k-1)N}[N-1]\,\Big([N]+1\Big).
}
\label{antiparlink}
\ee
This is the same as (\ref{antipvirt}) -- the first essentially new answer
for a virtual family in the present paper.

Like in the ordinary antiparallel case one can validate (or derive)
the quantization rule for $y_{n-2}$ by considering another coloring,
i.e. taking another vertex  of the hypercube. Another confirmation of the quantization rule is that it leads to the answer,
which is consistent with the evolution in $n$, i.e. has the form (\ref{antipevo})
with the $k$-independent coefficients $\gamma$ and $\delta$..

\section{Twist knot series \label{twist}}

\subsection{The four-vertex twist knot}

We begin from the first non-trivial example.
From the knot diagram with all black vertices and its resolution

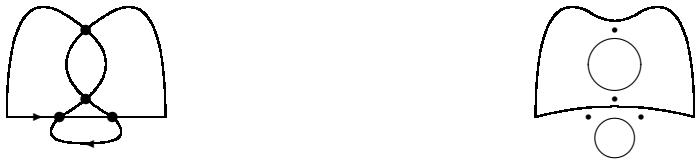
\begin{picture}(300,70)(-50,-20)
\put(0,0){\line(1,0){60}}
\qbezier(0,0)(0,50)(20,40)
\qbezier(60,0)(60,50)(40,40)
\qbezier(20,40)(55,20)(20,0)
\qbezier(40,40)(5,20)(40,0)
\qbezier(20,0)(10,-10)(30,-10)
\qbezier(40,0)(50,-10)(30,-10)
\put(20,0){\circle*{4}}
\put(40,0){\circle*{4}}
\put(30,7){\circle*{4}}
\put(30,33){\circle*{4}}
\put(12,0){\vector(1,0){2}}
\put(31,-10){\vector(-1,0){2}}
%
%
\qbezier(200,0)(230,8)(260,0)
\qbezier(200,0)(200,50)(220,40)
\qbezier(260,0)(260,50)(240,40)
\qbezier(220,40)(230,33)(240,40)
\put(230,20){\circle{20}}
\put(230,-8){\circle{15}}
\put(220,0){\circle*{2}}
\put(240,0){\circle*{2}}
\put(230,7){\circle*{2}}
\put(230,33){\circle*{2}}
\end{picture}

\noindent
we read the following primary and the main classical hypercubes:

\begin{equation}
\begin{array}{ccccc}
 & & 3 \\
& 2 & 1 & 2 \\
& 2 & 1 & 2 \\
3 &&&&1 \\
& 2 & 1 & 2 \\
& 2 & 1 & 2 \\
 & & 3
\end{array}
\ \ \ \longrightarrow \ \ \
\begin{array}{c|c|c|c|c}
 & & 2(N^3-N^2) \\
& N^3-N^2 & N(N-1)^2 & 2N(N-1)^2 \\
& N^3-N^2 & N(N-1)^2 & 2N(N-1)^2 \\
\boxed{N^3} &&&& \underline{(N-1)(3N-5)}  \\
& N^3-N^2 & N(N-1)^2 & 2N(N-1)^2 \\
& \boxed{\boxed{N^3-N^2}} & N(N-1)^2 & 2N(N-1)^2 \\
 & & \boxed{\boxed{\boxed{2(N^3-N^2)}}}
\end{array}
\label{twist2hypercubes}
\end{equation}
To avoid overloading the picture we do not show the edges -- but they are
implicitly used in the construction of the main hypercube.
The only non-trivial is the quantization of the underlined item --
and it can be obtained from the requirement that alternative coloring,

\begin{picture}(300,70)(-50,-20)
\put(0,0){\line(1,0){60}}
\qbezier(0,0)(0,50)(20,40)
\qbezier(60,0)(60,50)(40,40)
\qbezier(20,40)(55,20)(20,0)
\qbezier(40,40)(5,20)(40,0)
\qbezier(20,0)(10,-10)(30,-10)
\qbezier(40,0)(50,-10)(30,-10)
\put(20,0){\circle{4}}
\put(40,0){\circle*{4}}
\put(30,7){\circle*{4}}
\put(30,33){\circle*{4}}
\put(12,0){\vector(1,0){2}}
\put(31,-10){\vector(-1,0){2}}
\end{picture}

\noindent
which corresponds to picking the double-boxed vertex of the hypercube
as the initial one,
provides the unknot:
\begin{equation}
\begin{array}{r}
-q^{2N-3}[N]\left\{
[N][N-1]-q\Big([N]^2+2[N-1]^2+[2][N][N-1]\Big)
+ q^2\Big(3[N][N-1]+3[2][N-1]^2    \Big) - \right.  \\[0.3cm] \left.
-q^3\Big(2[N-1]^2+[2][N][N-1]+[N-1]\Big(\underline{[N-1]+[2][N-2]}\Big) \Big)
+q^4[2][N-1]^2
\right\} \ = \ [N].
\end{array}
\end{equation}

To obtain $[N]$ we had to take the underlined item to be what it is,
i.e. the relevant quantization rule in this case is
$\underline{(N-1)(3N-5)}\longrightarrow
[N-1]\Big(\underline{[N-1]+[2][N-2]}\Big)$
-- as already known from \cite{DM3}.

Using this prescription, we can obtain the HOMFLY polynomial for original
(all black) coloring, starting from the boxed main (Seifert) vertex of the
hypercube:

\begin{equation}
\begin{array}{c}
q^{4(N-1)}[N]\Big\{[N]^2-4q[N][N-1] + q^2\Big(2[2][N][N-1]+4[N-1]^2\Big)-
\\[0.2cm]
-4q^3[2][N-1]^2 +q^4[N-1]\Big(\underline{[N-1]+[2][N-2]}\Big)\Big\} =
\\[0.2cm]
= [N]\left\{\left(q^2+\cfrac{1}{q^2}\right)A^2 - A^4\right\}
= [N]\Big\{1-A^2\{Aq\}\{A/q\}\Big\}
\ = \  H^{3_1}_{_\Box},
\end{array}
\end{equation}
where $A=q^N$ and $\{Aq^{\pm 1}\} = (q-q^{-1})[N\pm 1] = q^{N\pm 1}- q^{-N\mp 1}$.

\bigskip

For another initial vertex,

\begin{picture}(300,70)(-50,-20)
\put(0,0){\line(1,0){60}}
\qbezier(0,0)(0,50)(20,40)
\qbezier(60,0)(60,50)(40,40)
\qbezier(20,40)(55,20)(20,0)
\qbezier(40,40)(5,20)(40,0)
\qbezier(20,0)(10,-10)(30,-10)
\qbezier(40,0)(50,-10)(30,-10)
\put(20,0){\circle{4}}
\put(40,0){\circle{4}}
\put(30,7){\circle*{4}}
\put(30,33){\circle*{4}}
\put(12,0){\vector(1,0){2}}
\put(31,-10){\vector(-1,0){2}}
\end{picture}

\noindent
marked by a triple box in (\ref{twist2hypercubes}), we get:

\begin{equation}
\begin{array}{r}
q^{-2}[N]\left\{
(1+q^4)[2][N][N-1] - (q+q^3)\Big(2[N][N-1]  + 2[2][N-1]^2\Big) + \right.
\\ [0.3cm]
\left. +q^2\left([N]^2+ 4[N-1]^2+[N-1]\Big(\underline{[N-1]+[2][N-2]}\Big)\right) \right\} =
\\[0.3cm]
= [N]\Big\{1+\{Aq\}\{A/q\}\Big\}
\ = \  H^{4_1}_{_\Box}.
\end{array}
\end{equation}

\subsection{The three-vertex virtual twist knot}

For the similar virtual knot

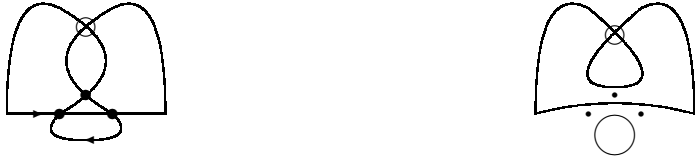
\begin{picture}(300,70)(-50,-20)
\put(0,0){\line(1,0){60}}
\qbezier(0,0)(0,50)(20,40)
\qbezier(60,0)(60,50)(40,40)
\qbezier(20,40)(55,20)(20,0)
\qbezier(40,40)(5,20)(40,0)
\qbezier(20,0)(10,-10)(30,-10)
\qbezier(40,0)(50,-10)(30,-10)
\put(20,0){\circle*{4}}
\put(40,0){\circle*{4}}
\put(30,7){\circle*{4}}
\put(30,33){\circle{7}}
\put(12,0){\vector(1,0){2}}
\put(31,-10){\vector(-1,0){2}}
%
%
\qbezier(200,0)(230,8)(260,0)
\qbezier(200,0)(200,50)(220,40)
\qbezier(260,0)(260,50)(240,40)
%
\put(230,-8){\circle{15}}
\qbezier(220,40)(255,10)(230,10)
\qbezier(240,40)(205,10)(230,10)
\put(220,0){\circle*{2}}
\put(240,0){\circle*{2}}
\put(230,7){\circle*{2}}
\put(230,30){\circle{7}}
\end{picture}

\noindent
the initial and full hypercubes are:
\begin{equation*}
\begin{array}{cccc}
& 1 & 2 \\
2 & 3 & 2 & 1 \\
& 1 & 2
\end{array}
\ \ \ \longrightarrow \ \ \
\begin{array}{c|c|c|c}
& N^2-N & -N^3+2N^2-N \\
\boxed{N^2} & N^2-N^3 & 2(N^2-N) & \underline{-N^3 +4N^2 - 3N} \\
& \boxed{\boxed{N^2-N}} & \boxed{\boxed{\boxed{ -N^3+2N^2-N}}}
\end{array}
\end{equation*}
Quantization of the underlined item is prescribed by the requirement that

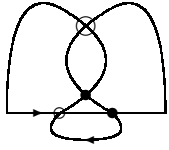
\begin{picture}(300,70)(-50,-20)
\put(0,0){\line(1,0){60}}
\qbezier(0,0)(0,50)(20,40)
\qbezier(60,0)(60,50)(40,40)
\qbezier(20,40)(55,20)(20,0)
\qbezier(40,40)(5,20)(40,0)
\qbezier(20,0)(10,-10)(30,-10)
\qbezier(40,0)(50,-10)(30,-10)
\put(20,0){\circle{4}}
\put(40,0){\circle*{4}}
\put(30,7){\circle*{4}}
\put(30,33){\circle{7}}
\put(12,0){\vector(1,0){2}}
\put(31,-10){\vector(-1,0){2}}
\end{picture}

\noindent
with the double-boxed initial vertex is the unknot:
\begin{equation}
\begin{array}{l}
V^{\stackrel{\bullet}{\circ\bullet}}_{_\Box} =
-q^{N-2}[N]\left\{[N-1] - q\Big([N]+[2][N-1]-[N-1]^2\Big) + \right.
\\[0.2cm]
 \left.
+q^2\Big( [N-1]-[N][N-1]-[N-1]\big(\underline{[N-2]-1}\big)   \Big)
+q^3[N-1]^2
\right\} \ = \ [N],
\end{array}
\end{equation}
i.e. the relevant quantization rule in this case is
$\underline{(N-1)(N-3)}\longrightarrow
[N-1]\big(\underline{[N-2]-1}\big)$  -- as already known from \cite{MMMvirt}.

Likewise the triple-boxed initial vertex, describing the coloring

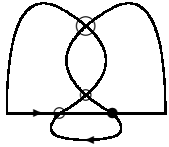
\begin{picture}(300,70)(-50,-20)
\put(0,0){\line(1,0){60}}
\qbezier(0,0)(0,50)(20,40)
\qbezier(60,0)(60,50)(40,40)
\qbezier(20,40)(55,20)(20,0)
\qbezier(40,40)(5,20)(40,0)
\qbezier(20,0)(10,-10)(30,-10)
\qbezier(40,0)(50,-10)(30,-10)
\put(20,0){\circle{4}}
\put(40,0){\circle*{4}}
\put(30,7){\circle{4}}
\put(30,33){\circle{7}}
\put(12,0){\vector(1,0){2}}
\put(31,-10){\vector(-1,0){2}}
\end{picture}

\noindent
also provides the unknot:
\begin{equation}
\begin{array}{r}
V^{\stackrel{\circ}{\circ\bullet}}_{_\Box} =
q^{-N-1}[N]\left\{-[N-1]^2 - q\Big([N-1] - [N][N-1] -[N-1]\Big([N-2]-1\Big)\Big) + \right.\\[0.2cm]
\left.
+q^2\Big([N]+[2][N-1]-[N-1]^2\Big) -q^3[N-1]\right\} = [N].
\end{array}
\end{equation}

Then the HOMFLY polynomial for original (all black) coloring (with the single-boxed
initial vertex) is
\begin{equation}
\begin{array}{c}
V^{\stackrel{\bullet}{\bullet\bullet}}_{_\Box} =
q^{3(N-1)}[N]\left\{
[N] - q\Big(2[N-1]-[N][N-1]\Big) + q^2\big([2][N-1]-2[N-1]^2\big)+\right.
\\[0.3cm]
\left.+q^3[N-1]\Big(\underline{[N-2]-1}\big)
\right\} = [N]\left\{q^{N+1}+q^{2N-2}-q^{3N-1} \right\} = V^{2.1}_{\Box} =
\\[0.3cm]
= [N]\Big(1 -q^{N-1}(q^{N+1}-1)(q^{N-1}-q^{1-N})\Big) =
[N]\left(1-\frac{A}{q}(Aq-1)\{A/q\}\right).
\end{array}
\label{2pbbb}
\end{equation}

For yet another coloring,

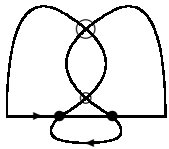
\begin{picture}(300,70)(-50,-20)
\put(0,0){\line(1,0){60}}
\qbezier(0,0)(0,50)(20,40)
\qbezier(60,0)(60,50)(40,40)
\qbezier(20,40)(55,20)(20,0)
\qbezier(40,40)(5,20)(40,0)
\qbezier(20,0)(10,-10)(30,-10)
\qbezier(40,0)(50,-10)(30,-10)
\put(20,0){\circle*{4}}
\put(40,0){\circle*{4}}
\put(30,7){\circle{4}}
\put(30,33){\circle{7}}
\put(12,0){\vector(1,0){2}}
\put(31,-10){\vector(-1,0){2}}
\end{picture}

\noindent
initial is the double boxed vertex in
\begin{equation*}
\begin{array}{c|c|c|c}
& [N][N-1] & -[N][N-1]^2 \\  && \\
\boxed{[N]^2} & \boxed{\boxed{-[N]^2[N-1]}} & \boxed{\boxed{\boxed{[2][N][N-1]}}}
&  -[N][N-1]\Big([N-2]-1\Big) \\  && \\
&  [N][N-1]  &  -[N][N-1]^2
\end{array}
\end{equation*}
and
\begin{equation}
\begin{array}{l}
V^{\stackrel{\circ}{\bullet\bullet}}_{_\Box} =
-q^{N-2}[N]\left\{-[N][N-1]-q\Big([N]-2[N-1]^2\Big)+q^2\Big(2[N-1]-[N-1]\big([N-2]-1\big)\Big)\right.-
\\[0.3cm]
\left.-q^3[2][N-1]
\right\} = [N]\Big(1 +  (q^{N+1}-1)(q^{N-1}-q^{1-N})   \Big)
= [N]\Big(1 +  (Aq-1)\{A/q\}\Big) = V^{3.2}_{\Box},
\end{array}
\label{2pwbb}
\end{equation}
while for the triple-boxed initial vertex,

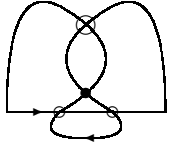
\begin{picture}(300,70)(-50,-20)
\put(0,0){\line(1,0){60}}
\qbezier(0,0)(0,50)(20,40)
\qbezier(60,0)(60,50)(40,40)
\qbezier(20,40)(55,20)(20,0)
\qbezier(40,40)(5,20)(40,0)
\qbezier(20,0)(10,-10)(30,-10)
\qbezier(40,0)(50,-10)(30,-10)
\put(20,0){\circle{4}}
\put(40,0){\circle{4}}
\put(30,7){\circle*{4}}
\put(30,33){\circle{7}}
\put(12,0){\vector(1,0){2}}
\put(31,-10){\vector(-1,0){2}}
\end{picture}

\noindent
we would  get:
\be
V^{\stackrel{\bullet}{\circ\circ}}_{_\Box} =
q^{-N-1}[N]\left\{[2][N-1] - q\Big(2[N-1]-[N-1]\big(\underline{[N-2]-1}]\big)\Big)
+ q^2\Big([N] - 2[N-1]^2\Big) +q^3[N][N-1]\right\} =
\nn
\ee
\vspace{-0.3cm}
\be
= [N]\Big(1 + q^{-N-1}(q^{N+1}-1)(q^{N-1}-q^{1-N})\Big)
= [N]\left(1+\frac{1}{Aq}(Aq-1)\{A/q\}\right)
= V^{\stackrel{\circ}{\bullet\bullet}}_{_\Box}(q^{-1},A^{-1}).
\label{2pbww}
\ee

Finally, for the last coloring

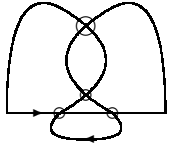
\begin{picture}(300,70)(-50,-20)
\put(0,0){\line(1,0){60}}
\qbezier(0,0)(0,50)(20,40)
\qbezier(60,0)(60,50)(40,40)
\qbezier(20,40)(55,20)(20,0)
\qbezier(40,40)(5,20)(40,0)
\qbezier(20,0)(10,-10)(30,-10)
\qbezier(40,0)(50,-10)(30,-10)
\put(20,0){\circle{4}}
\put(40,0){\circle{4}}
\put(30,7){\circle{4}}
\put(30,33){\circle{7}}
\put(12,0){\vector(1,0){2}}
\put(31,-10){\vector(-1,0){2}}
\end{picture}

\noindent
we can get the answer directly from (\ref{2pbbb}):
\begin{equation}
V^{\stackrel{\circ}{\circ\circ}}_{_\Box}(q,A) =
V^{\stackrel{\bullet}{\bullet\bullet}}_{_\Box}(q^{-1},A^{-1}) =
[N]\left(1-\frac{1}{A^2}(Aq-1)\{A/q\}\right).
\label{2pwww}
\end{equation}

\subsection{Evolution for ordinary and virtual twist knots}

Consideration of the single knot diagram in the previous section
is actually enough to obtain the answer for entire family of twist knots
-- such is the power of the evolution method.
As mentioned in \ref{antipsec}, for antiparallel strands evolution
actually changes the number of vertices by two, however in the case
of twist knots there is an additional topological identity

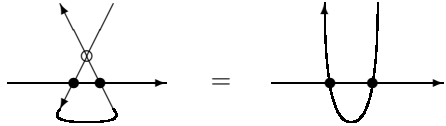
\begin{picture}(300,70)(-150,-30)
\put(-15,0){\vector(1,0){60}}
\put(25,-10){\vector(-1,2){20}}
\put(25,30){\vector(-1,-2){20}}
\put(10,0){\circle*{4}}
\put(20,0){\circle*{4}}
\put(15,10){\circle{4}}
\qbezier(5,-10)(0,-15)(15,-15)
\qbezier(25,-10)(30,-15)(15,-15)
\put(62,-2){\mbox{$=$}}
\put(85,0){\vector(1,0){65}}
\put(107,0){\circle*{4}}
\put(123,0){\circle*{4}}
\qbezier(105,30)(105,-15)(115,-15)
\qbezier(125,30)(125,-15)(115,-15)
\put(105,28){\vector(0,1){2}}

\end{picture}

\noindent
which allows one to relate the two series -- with even and odd number of vertices:
\begin{equation}
V^{\stackrel{2k+1}{\bullet\bullet}} = V^{\stackrel{2k}{\bullet\bullet}},
\label{V-V}
\end{equation}
where $2k+1$ is the number of black vertices on the vertical (sterile crossing not counted).
This is the analogue of the same relation holds for the ordinary twist knots  \cite{evo,DM3}:
\begin{equation}\boxed{
H^{\stackrel{2k}{\bullet\bullet}}_{_\Box} = H^{\stackrel{2k-1}{\bullet\bullet}}_{_\Box} =
1 - \frac{A^{k+1}\{A^k\}}{\{A\}}\,\{Aq\}\{A/q\},
}
\label{H-H}
\end{equation}
with the following knots being different members of the series:
\begin{equation}
\begin{array}{rcl}
& \ldots & \\
5_2 & k=2 &  \stackrel{3}{\bullet\bullet} = \stackrel{4}{\bullet\bullet} \\
3_1 & k=1 &  \stackrel{1}{\bullet\bullet} = \stackrel{2}{\bullet\bullet} \\
unknot & k = 0 & \stackrel{-1}{\bullet\bullet} = \stackrel{0}{\bullet\bullet}  \\
4_1 & k =-1 & \stackrel{-3}{\bullet\bullet} = \stackrel{-2}{\bullet\bullet} \\
& \ldots &
\end{array}
\end{equation}
Note only that in our notation $V^{\stackrel{2k+1}{\bullet\bullet}}$ is similar
to $H^{\stackrel{2k+2}{\bullet\bullet}}$ and therefore relation (\ref{V-V})
is somewhat different from that in (\ref{H-H}).

From (\ref{2pbbb}) and (\ref{2pwbb}) with $2k+1=1$ and $2k+1=-1$ respectively
we can get the coefficients $\gamma$ and $\delta$ in the evolution formula (\ref{antipevo}):
\begin{equation}
\begin{array}{c}
V^{\stackrel{2k+1}{\bullet\bullet}}_{_\Box}=
[N]\left\{1\ + \ \boxed{(Aq-1)\{A/q\}}\left(1  - \left(1+\frac{A}{q}\right)
\frac{A^{2k+2}-1}{A^2-1}\right)\right\} =
\\[0.3cm]
= [N] \left( 1\ +\ \frac{A\{Aq\}\{A/q\}}{\{A\}} \
-\ \frac{(A+q)(Aq-1)\{A/q\}}{q\{A\}}\cdot A^{2k+1}\right) =
\\[0.3cm]
\boxed{=[N]\left(1 \ +\ \frac{ (Aq+1)-(A+q)A^{2k+1} }{q\{A\}} (Aq-1)\{A/q\}\right).}
\end{array}
\label{Vbb}
\end{equation}
Similarly from  (\ref{2pbww}) and (\ref{2pwww}) we deduce:
\begin{equation}
\begin{array}{c}
V^{\stackrel{2k+1}{\circ\circ}}_{_\Box} =
[N] \left(-\frac{1}{A^2}+ \frac{A^{2k+2}-1}{A^2-1}\left(\frac{1}{Aq} + \frac{1}{A^2}\right)
\right)=
\\[0.3cm]
= [N]\left(1 \ -\ \frac{ (Aq+1)-(A+q)A^{2k+1} }{qA^2\{A\}} (Aq-1)\{A/q\}\right).
\end{array}
\label{Vww}
\end{equation}
In fact, these two formulae are not independent:
\begin{equation}
V^{\stackrel{2k+1}{\circ\circ}}_{_\Box}(q,A) =
V^{\stackrel{-2k-1}{\bullet\bullet}}_{_\Box}(q^{-1},A^{-1}).
\label{bbVww}
\end{equation}
Together with (\ref{V-V}) eq.(\ref{Vbb}) provides the full answer for
the family of twist virtual knots.
In particular,
\begin{equation}
\begin{array}{l}
\ldots
\\[0.3cm]
G^{\stackrel{-6}{\bullet\bullet}}_{_\Box}=
G^{\stackrel{-5}{\bullet\bullet}}_{_\Box}=
1+\frac{1}{qA}+\frac{1}{A^2}+\frac{1}{qA^3}+\frac{1}{A^4},
\\[0.2cm]
G^{\stackrel{-4}{\bullet\bullet}}_{_\Box}=
G^{\stackrel{-3}{\bullet\bullet}}_{_\Box}=
1+\frac{1}{qA}+\frac{1}{A^2},
\\[0.2cm]
G^{\stackrel{-2}{\bullet\bullet}}_{_\Box}=
\boxed{G^{\stackrel{-1}{\bullet\bullet}}_{_\Box}=
G^{\stackrel{\circ}{\bullet\bullet}}_{_\Box}}= 1,
\\[0.2cm]
G^{\stackrel{0}{\bullet\bullet}}_{_\Box}=
\boxed{G^{\stackrel{1}{\bullet\bullet}}_{_\Box} =
G^{\stackrel{\bullet}{\bullet\bullet}}_{_\Box}}=-\frac{A}{q},
\\[0.2cm]
G^{\stackrel{2}{\bullet\bullet}}_{_\Box}=
G^{\stackrel{3}{\bullet\bullet}}_{_\Box}=
-\frac{A}{q}(1+qA+A^2),
\\[0.2cm]
G^{\stackrel{4}{\bullet\bullet}}_{_\Box}=
G^{\stackrel{5}{\bullet\bullet}}_{_\Box}=
-\frac{A}{q}(1+qA+A^2+qA^3+A^4),
\\
\ldots \\[0.6cm]
G^{\stackrel{-3}{\circ\circ}}_{_\Box}
 = -\frac{1}{A^2}\left(1+\frac{1}{qA}+\frac{1}{A^2}\right)
= G^{\stackrel{-2}{\circ\circ}}_{_\Box},
\\[0.2cm]
\boxed{G^{\stackrel{-1}{\circ\circ}}_{_\Box}
= G^{\stackrel{\circ}{\circ\circ}}_{_\Box}}= -\frac{1}{A^2}
= G^{\stackrel{0}{\circ\circ}}_{_\Box},
\\[0.2cm]
\boxed{G^{\stackrel{1}{\circ\circ}}_{_\Box}
= G^{\stackrel{\bullet}{\circ\circ}}_{_\Box}}=\frac{1}{qA}
= G^{\stackrel{2}{\circ\circ}}_{_\Box},
\\[0.2cm]
G^{\stackrel{3}{\circ\circ}}_{_\Box}
 = \frac{1+qA+A^2}{qA}
= G^{\stackrel{4}{\circ\circ}}_{_\Box},
\\
\ldots
\end{array}
\label{predtwist}
\end{equation}
where listed are the coefficients $G$ in
\begin{equation}
\boxed{
V_{_\Box}^{\cal L}(A.q) = [N]\Big( 1 + G^{\cal L}_{_\Box}(A,q) \cdot (Aq-1)\{A/q\}\Big).
}
\label{difvirtfund}
\end{equation}
From (\ref{bbVww}) it follows that
$
\ \ qA \cdot G^{\stackrel{2k+1}{\circ\circ}}_{_\Box}(q,A) =
G^{\stackrel{-2k-1}{\bullet\bullet}}_{_\Box}(q^{-1},A^{-1}) \ \
$
because of the change of the factor $(Aq-1)$.

\bigskip

In the next section \ref{twistcheck} we check the prediction (\ref{predtwist})
of evolution method by explicit calculation of a few additional examples.
The four quantities that we already calculated -- and used to make this prediction --
are put in boxes.

\section{More examples of (virtual) twist knots \label{twistcheck}}

\subsection{The six-vertex twist knot ($2k+1=\pm 3$)}

Switching to more complicated examples we start from direct generalization
of what we just studied:

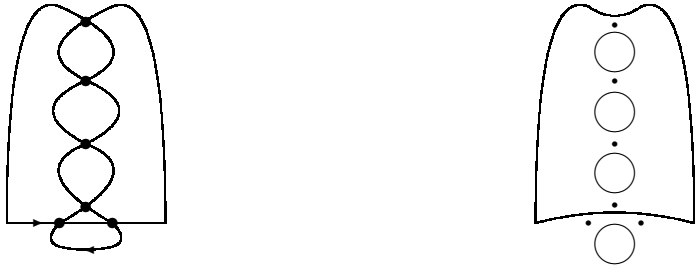
\begin{picture}(300,110)(-50,-20)
\put(0,0){\line(1,0){60}}
\qbezier(0,0)(0,90)(20,82)
\qbezier(60,0)(60,90)(40,82)
\qbezier(30,30)(55,20)(20,0)
\qbezier(30,30)(5,20)(40,0)
\qbezier(30,30)(55,43)(30,54)
\qbezier(30,30)(5,43)(30,54)
\qbezier(20,82)(55,66)(30,54)
\qbezier(40,82)(5,66)(30,54)
\qbezier(20,0)(10,-10)(30,-10)
\qbezier(40,0)(50,-10)(30,-10)
\put(20,0){\circle*{4}}
\put(40,0){\circle*{4}}
\put(30,6){\circle*{4}}
\put(30,30){\circle*{4}}
\put(30,54){\circle*{4}}
\put(30,76){\circle*{4}}
\put(12,0){\vector(1,0){2}}
\put(31,-10){\vector(-1,0){2}}
%
%
\qbezier(200,0)(230,8)(260,0)
\qbezier(200,0)(200,90)(220,82)
\qbezier(260,0)(260,90)(240,82)
\qbezier(220,82)(230,75)(240,82)
\put(230,65){\circle{15}}
\put(230,42){\circle{15}}
\put(230,19){\circle{15}}
\put(230,-8){\circle{15}}
\put(220,0){\circle*{2}}
\put(240,0){\circle*{2}}
\put(230,7){\circle*{2}}
\put(230,30){\circle*{2}}
\put(230,54){\circle*{2}}
\put(230,75){\circle*{2}}
\end{picture}

\noindent
We temporarily omitted the 3-point case, because
resolved diagrams there would have a different structure
-- see s.\ref{twist3p} below.

The primary hypercube is now made from the following vertices
(underlined are the cycle numbers, and the numbers in front of them
are their multiplicities):
\begin{equation*}
\setlength{\arraycolsep}{2pt}
\begin{array}{|ccccc|}
\underline{5} && 2\times \boxed{\boxed{\underline{4}}} && {\underline{5}} \\ &&&& \\
4\times\underline{4} && 8\times \underline{3} && 4\times\underline{4} \\ &&&& \\
6\times\underline{3} && 12\times \underline{2} && 6\times\underline{3} \\ &&&& \\
4\times\underline{2} && 8\times \underline{1} && 4\times\underline{2} \\ &&&& \\
\underline{3} && 2\times \underline{2} && \underline{1} \\ &&&& \\
\end{array}
\ \ \ \ \ \longrightarrow \ \ \ \ \
\begin{array}{|ccccc|}
\boxed{[N]^5} && 2\times\boxed{\boxed{[N]^4[N-1]}} && [2][N]^4[N-1]\\ &&&&\\
4\times[N]^4[N-1]&& 8\times [N]^3[N-1]^2 && 4\times [2][N]^3[N-1]^2 \\ &&&&\\
6\times [N]^3[N-1]^2 && 12\times [N]^2[N-1]^3 && 6\times [2][N]^2[N-1]^3 \\ &&&&\\
4\times [N]^2[N-1]^3 && 8\times [N][N-1]^4 && 4\times[2][N][N-1]^4 \\ &&&&\\
\l[N]^2[N-1]Y_3 && 2\times[N][N-1]^2Y_3 && [N][N-1]\Big([N-1]^3+[N-2]Y_3\Big)
\end{array},
\setlength{\arraycolsep}{6pt}
\end{equation*}
where $Y_3=[N-1][N-2]+[2]$ is the quantization of $N^2-3N+4$, obtained
from the requirement that we get the unknot, if we take for initial the
double boxed vertex, see s.5.7.4 of \cite{DM3}.

\bigskip

For the virtual knot with a similar knot diagram

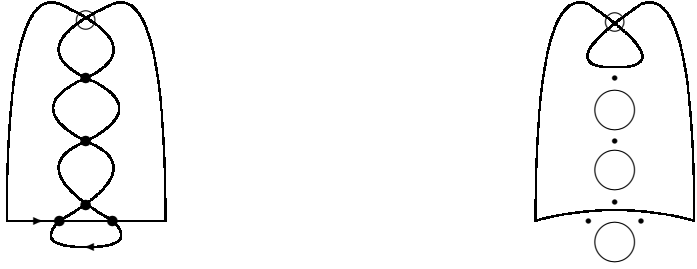
\begin{picture}(300,110)(-50,-20)
\put(0,0){\line(1,0){60}}
\qbezier(0,0)(0,90)(20,82)
\qbezier(60,0)(60,90)(40,82)
\qbezier(30,30)(55,20)(20,0)
\qbezier(30,30)(5,20)(40,0)
\qbezier(30,30)(55,43)(30,54)
\qbezier(30,30)(5,43)(30,54)
\qbezier(20,82)(55,66)(30,54)
\qbezier(40,82)(5,66)(30,54)
\qbezier(20,0)(10,-10)(30,-10)
\qbezier(40,0)(50,-10)(30,-10)
\put(20,0){\circle*{4}}
\put(40,0){\circle*{4}}
\put(30,6){\circle*{4}}
\put(30,30){\circle*{4}}
\put(30,54){\circle*{4}}
\put(30,76){\circle{7}}
\put(12,0){\vector(1,0){2}}
\put(31,-10){\vector(-1,0){2}}
%
%
\qbezier(200,0)(230,8)(260,0)
\qbezier(200,0)(200,90)(220,82)
\qbezier(260,0)(260,90)(240,82)
\qbezier(220,82)(255,58)(230,58)
\qbezier(240,82)(205,58)(230,58)
%
\put(230,42){\circle{15}}
\put(230,19){\circle{15}}
\put(230,-8){\circle{15}}
\put(220,0){\circle*{2}}
\put(240,0){\circle*{2}}
\put(230,7){\circle*{2}}
\put(230,30){\circle*{2}}
\put(230,54){\circle*{2}}
\put(230,75){\circle{7}}
\end{picture}

\noindent
we get the following primary and main hypercubes:

\bigskip

$
\setlength{\arraycolsep}{1pt}
\begin{array}{ccccc}
\underline{4}&& 2\times\boxed{\boxed{\underline{3}}} && \boxed{\underline{4}} \\ \\
3\times\underline{3} && 6\times\underline{2} && 3\times\underline{3} \\ \\
3\times\underline{2} && 6\times\underline{1} && 3\times\underline{2} \\ \\
\underline{3} && 2\times \underline{2} && \underline{1}
\end{array}
\ \ \ \ \ \longrightarrow \ \ \ \ \
\begin{array}{ccccc}
\boxed{[N]^4} && 2\times \boxed{\boxed{[N]^3[N-1]}} && \boxed{\boxed{\boxed{[2][N]^3[N-1]}}}\\
&&&&\\
3\times [N]^3[N-1] && 6\times [N]^2[N-1]^2 && 3\times [2][N]^2[N-1]^2 \\ &&&& \\
3\times [N]^2[N-1]^2 && 6 \times [N][N-1]^3 && 3\times [2][N][N-1]^3 \\ &&&& \\
\l[N]^2[N-1]\Big(\underline{[N-2]-1}\Big)
&& 2\times [N][N-1]^2\Big(\underline{[N-2]-1}\Big) && [N][N-1]\underline{y_2}
\end{array}
\setlength{\arraycolsep}{6pt}
$

\bigskip

\noindent
Starting from the double boxed vertex, we should get the unknot and
this defines the quantization rule for the underlined items in last row,
in particular, $2N^2-7N+7 \longrightarrow y_2 = [2][N-1][N-2]-[N-2]+1$.
Indeed
\begin{equation}
\begin{array}{c}
-q^{3N-4}[N]\left\{[N]^2[N-1] - q\Big([N]^3+3[N][N-1]^2+[2][N]^2[N-1]\Big)+\right.
\\[0.2cm]
+q^2\Big(3[N]^2[N-1]+[N]^2[N-1]+3[2][N][N-1]^2+3[N-1]^3\Big) -
\\[0.2cm]
- q^3\Big(3[N][N-1]^2 +[N-1]^2\big([N-2]-1\big)+3[N][N-1]^2 +3[2][N-1]^3  \Big)
\\[0.2cm]
+q^4\Big( 3[N-1]^3+[N][N-1]\big([N-2]-1\big) + [N-1]y_2 \Big) -
\\[0.2cm]
\left.-q^5 [N-1]^2\Big([N-2]-1\Big)
\right\} = [N].
\end{array}
\end{equation}

Now we can find HOMFLY polynomial for the original diagram (all vertices black):
\begin{equation}
\begin{array}{c}
q^{5N-5}[N]\left\{[N]^3-5q[N]^2[N-1]+q^2\Big(9[N][N-1]^2+[2][N]^2[N-1]\Big) - \right.
\\[0.2cm]
-q^3\Big([N][N-1]\big([N-2]-1\big) + 6[N-1]^3+3[2][N][N-1]^2\Big)+
\\[0.2cm]
+ q^4\Big(2[N-1]^2\big([N-2]-1)+3[2][N-1]^3\Big) -
\\[0.2cm]
\left.-q^5[N-1]\Big([2][N-1][N-2]-[N-2]+1\Big)\right\}
= [N]\Big(1 - q^{N-1}(q^{N+1}-1)\{A/q\}(q^{2N}+q^{N+1}+1)\Big).
\end{array}
\label{4pbb}
\end{equation}

If the triple boxed vertex is taken for initial one we get:
\begin{equation}
\begin{array}{c}
q^{N-3}[N]\left\{ [2][N]^2[N-1] - q\Big(2[N]^2[N-1]+3[2][N][N-1]^2\Big)\right.+
\\[0.2cm]
+ q^2\Big([N]^3+6[N][N-1]^2+3[2][N-1]^3\Big) -
 q^3\Big(3[N]^2[N-1]+6[N-1]^3+[N-1]y_2\Big)+
\\[0.2cm]
\left.
+q^4\Big(3[N][N-1]^2+2[N-1]^2\big([N-2]-1\big)\Big)
-q^5[N][N-1]\big([N-2]-1\big)\right\}\ =
\\[0.2cm]
= [N]\Big(1 + q^{-N-1}(q^{N+1}-1)\{A/q\}(q^{2N}+q^{N+1}+1)\Big).
\end{array}
\label{4pbw}
\end{equation}

\subsection{The three-vertex twist knot ($2k=0$)
\label{twist3p}}

This example would be of course the first one. We did not begin from it, because it provides just one point on the evolution line and would not be sufficient to restore the evolution for $ virtual$  knots - thus we began from the four-vertex example with two virtual points $2k+1=\pm 1$. Once we did that, the case $2k+1=\pm 3 $ was the natural next example, because the structure of Seifert cycles differs drastically depending on the parity of the points number. Still we would like to look at this second series also.

From the knot diagram with all black vertices and its resolutions

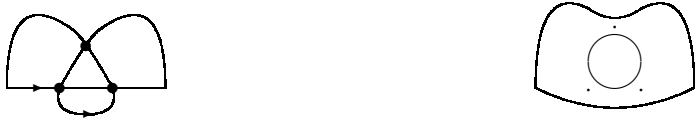
\begin{picture}(300,60)(-50,-20)
\put(0,0){\line(1,0){60}}
\qbezier(0,0)(0,40)(25,22)
\qbezier(60,0)(60,40)(35,22)
\qbezier(25,22)(30,18)(40,0)
\qbezier(35,22)(30,18)(20,0)

\qbezier(20,0)(17,-10)(30,-10)
\qbezier(40,0)(43,-10)(30,-10)
\put(20,0){\circle*{4}}
\put(40,0){\circle*{4}}
\put(30,16){\circle*{4}}

\put(12,0){\vector(1,0){2}}
\put(31,-10){\vector(1,0){2}}
\qbezier(260,0)(230,-15)(200,0)
\qbezier(260,0)(260,40)(240,30)
\qbezier(200,0)(200,40)(220,30)
\qbezier(220,30)(230,23)(240,30)
\put(230,10){\circle{20}}
\put(230,23){\circle*{1}}
\put(220,-1){\circle*{1}}
\put(240,-1){\circle*{1}}

\end{picture}

\noindent we read the following primary and the main classical hypercubes:
\begin{equation}
\begin{array}{cccc}
& 1 & 2\\
2 & 1 & 2 & 1\\
& 1 & 2\\
\end{array}
\longrightarrow
\label{tref}
\begin{array}{c|c|c|c}
& [N][N-1] & [2][N][N-1] \\
\boxed{[N]^2} & \boxed{ \boxed{[N][N-1]}}& [2][N][N-1] & [2]^2 [N][N-1] \\
&  [N][N-1]& [2][N][N-1]\\
\end{array}
\end{equation}
The HOMFLY polynomial for original (all black) coloring (with the boxed initial vertex) is

\begin{equation}
\begin{aligned}
q^{3(N-1)}[N] \Big ([N]-3q[N-1]+3q^2[2][N-1]-q^3[2]^2[N-1] \Big )= \\
= [N] \left ( 1-\frac{1}{q^2}(Aq-1)(Aq+1)(A-q)(A+q) \right ) =\\
=[N] \left ( 1-A^2  \left \{Aq \right \} \left \{A/q \right \} \right ).
\label{tp}
\end{aligned}
\end{equation}

For another initial vertex

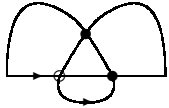
\begin{picture}(300,60)(-50,-20)
\put(0,0){\line(1,0){60}}
\qbezier(0,0)(0,40)(25,22)
\qbezier(60,0)(60,40)(35,22)
\qbezier(25,22)(30,18)(40,0)
\qbezier(35,22)(30,18)(20,0)

\qbezier(20,0)(17,-10)(30,-10)
\qbezier(40,0)(43,-10)(30,-10)
\put(20,0){\circle{4}}
\put(40,0){\circle*{4}}
\put(30,16){\circle*{4}}

\put(12,0){\vector(1,0){2}}
\put(31,-10){\vector(1,0){2}}
\end{picture}

\noindent marked by a double box in (\ref{tref}), we naturally get the answer for the unknot:
\begin{equation}
\begin{aligned}
-q^{2N-1}[N] \Big ([N-1]-q \left ([N]+2[2][N-1] \right )+q^2 \left (2[N-1]+[2]^2[N-1] \right )-q^3[2][N-1] \Big )= [N].
\end{aligned}
\end{equation}

\subsection{The two-vertex virtual twist knot}

\noindent
For the similar virtual knot

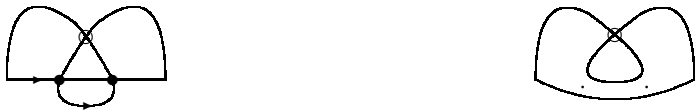
\begin{picture}(300,60)(-50,-20)
\put(0,0){\line(1,0){60}}
\qbezier(0,0)(0,40)(25,22)
\qbezier(60,0)(60,40)(35,22)
\qbezier(25,22)(30,18)(40,0)
\qbezier(35,22)(30,18)(20,0)

\qbezier(20,0)(17,-10)(30,-10)
\qbezier(40,0)(43,-10)(30,-10)
\put(20,0){\circle*{4}}
\put(40,0){\circle*{4}}
\put(30,16){\circle{5}}

\put(12,0){\vector(1,0){2}}
\put(31,-10){\vector(1,0){2}}
\qbezier(260,0)(230,-15)(200,0)
\qbezier(260,0)(260,35)(240,25)
\qbezier(200,0)(200,35)(220,25)
\qbezier(240,25)(205,-1)(230,-1)
\qbezier(220,25)(255,-1)(230,-1)

\put(230,17){\circle{5}}
\put(218,-3){\circle*{1}}
\put(242,-3){\circle*{1}}

\end{picture}

\noindent  primary and the main classical hypercubes are:
\begin{equation}
\begin{array}{ccc}
&2\\
1 && 1\\
& 2\\
\end{array}
\label{2.1}
\longrightarrow
\begin{array}{c|c|c}
& [N][1-N] \\
\boxed{[N]} & & [2][N][1-N] \\
&\boxed{ \boxed{ [N][1-N] }}
\end{array}
\end{equation}
Fortunately, we can easily quantize all items in this example and obtain the HOMFLY polynomial for original (all black) coloring, starting from the boxed vertex of the hypercube:

\begin{equation}
\begin{aligned}
q^{2(N-1)}[N] \Big (1-2q[1-N]+q^2[2][1-N] \Big )=[N] \left (1- \frac{A}{q} \left (Aq-1 \right ) \left\{  A/q \right\} \right ).
\end{aligned}
\end{equation}
For another coloring of the knot with initial double-boxed vertex HOMFLY polynomial is

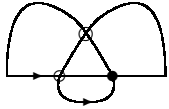
\begin{picture}(300,60)(-50,-20)
\put(0,0){\line(1,0){60}}
\qbezier(0,0)(0,40)(25,22)
\qbezier(60,0)(60,40)(35,22)
\qbezier(25,22)(30,18)(40,0)
\qbezier(35,22)(30,18)(20,0)

\qbezier(20,0)(17,-10)(30,-10)
\qbezier(40,0)(43,-10)(30,-10)
\put(20,0){\circle{4}}
\put(40,0){\circle*{4}}
\put(30,16){\circle{5}}

\put(12,0){\vector(1,0){2}}
\put(31,-10){\vector(1,0){2}}
\end{picture}

\begin{equation}
\begin{aligned}
-q^{-1}[N] \Big ([1-N]-q \left (1+[2][1-N] \right )+q^2[1-N] \Big )=[N].
\end{aligned}
\end{equation}

\subsection{The five-vertex twist knot $(2k=\pm 2)$ }

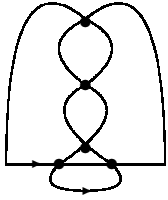
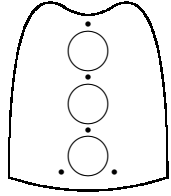
\begin{picture}(300,90)(-50,-20)
\put(0,0){\line(1,0){60}}
\qbezier(0,0)(0,80)(30,55)
\qbezier(60,0)(60,80)(30,55)
\qbezier(30,30)(50,20)(20,0)
\qbezier(30,30)(10,20)(40,0)
\qbezier(30,30)(50,45)(30,55)
\qbezier(30,30)(10,45)(30,55)

\qbezier(20,0)(10,-10)(30,-10)
\qbezier(40,0)(50,-10)(30,-10)
\put(20,0){\circle*{4}}
\put(40,0){\circle*{4}}
\put(30,6){\circle*{4}}
\put(30,30){\circle*{4}}
\put(30,54){\circle*{4}}

\put(12,0){\vector(1,0){2}}
\put(31,-10){\vector(1,0){2}}
\qbezier(260,0)(230,-10)(200,0)
\qbezier(260,0)(260,75)(240,65)
\qbezier(200,0)(200,75)(220,65)
\qbezier(240,65)(230,58)(220,65)
\put(230,8){\circle{15}}
\put(230,28){\circle{15}}
\put(230,48){\circle{15}}
\put(230,58){\circle*{2}}
\put(230,18){\circle*{2}}

\put(230,38){\circle*{2}}
\put(220,2){\circle*{2}}
\put(240,2){\circle*{2}}

\end{picture}

\noindent
The primary hypercube has the strucrure:
\begin{equation*}
\begin{array}{cccccc}
 & & 2 & 1 \\
 & & 2 & 1 \\
& 3 &2 &1&2\\
& 3 & 2 &{3} &{2}\\
4 & 3 & {4} & {3} & {2} & {1} \\
& 3 & 2 &{3} &{2} \\
& 3 & 2 & 1 & 2\\
 & & 2 & 1 \\
 & & 2 & 1 \\
 & & 2 & 1 \\
\end{array}
\end{equation*}
and the main classical hypercube is
\begin{equation*}
\begin{array}{c|c|c|c|c|c}
 & & N^2(N-1)^2 & N(N-1)^3 \\
 & & N^2(N-1)^2 & N(N-1)^3 \\
& \boxed { \boxed{ N^3(N-1)}} & N^2(N-1)^2 & N(N-1)^3 & \underline{N(N-1)(N^2-3N+4)}\\
& N^3(N-1) & N^2(N-1)^2 & 2N^2(N-1)^2 & 2N(N-1)^3\\
\boxed{N^4} &  N^3(N-1) & 2N^3(N-1) & 2N^2(N-1)^2 & 2N(N-1)^3 & \underline{2N(N-1)(N^2-3N+4)}  \\
& N^3(N-1) & N^2(N-1)^2 & 2N^2(N-1)^2 & 2N(N-1)^3\\
& N^3(N-1) & N^2(N-1)^2 & N(N-1)^3 & \underline{N(N-1)(N^2-3N+4)}\\
 & & N^2(N-1)^2 & N(N-1)^3 \\
 & & N^2(N-1)^2 & N(N-1)^3 \\
 & & N^2(N-1)^2 & N(N-1)^3 \\
\end{array}
\end{equation*}

We can predict quantization of the underlined item by the requirement that the coloring with the double-boxed initial vertex is the trefoil, polynomial of which is already known from (\ref{tp}):

\begin{picture}(300,90)(-50,-20)
\put(0,0){\line(1,0){60}}
\qbezier(0,0)(0,80)(30,55)
\qbezier(60,0)(60,80)(30,55)
\qbezier(30,30)(50,20)(20,0)
\qbezier(30,30)(10,20)(40,0)
\qbezier(30,30)(50,45)(30,55)
\qbezier(30,30)(10,45)(30,55)

\qbezier(20,0)(10,-10)(30,-10)
\qbezier(40,0)(50,-10)(30,-10)
\put(20,0){\circle*{4}}
\put(40,0){\circle*{4}}
\put(30,7){\circle*{4}}
\put(30,30){\circle*{4}}
\put(30,55){\circle{4}}

\put(12,0){\vector(1,0){2}}
\put(31,-10){\vector(1,0){2}}
\end{picture}

\begin{equation}
\begin{aligned}
q^{3N-4}[N] \Big ([N]^2[N-1]-q([N]^3+4[N][N-1]^2)+q^2(4[N]^2[N-1]+5[N-1]^3+[2][N][N-1]^2)-\\
-q^3(5[N][N-1]^2+[2][N]^2[N-1]+2[N-1]([N-1][N-2]+[2])+2[2][N-3]^3)+\\
+q^4(2[N-1]^3+2[2][N][N-1]^2+[2][N-1]([N-1][N-2]+[2]))-q^5[2][N-1]^3 \Big )=\\
=[N] \left ( 1-A^2  \left \{Aq \right \} \left \{A/q \right \} \right ),
\end{aligned}
\end{equation}
i.e. the relevant quantization rule in this case is $ \underline{ (N-1)(N^2-3N+4)} \rightarrow [N-1] \left ([N-1][N-2]+[2] \right ) $. Now we can get HOMFLY polynomial for original (all black) diagram:
\begin{equation}
\begin{aligned}
q^{5N-5}[N] \Big ( [N]^3-5q[N]^2[N-1]+q^2(9[N][N-1]^2+[2][N]^2[N-1])-q^3(3[2][N][N-1]^2+7[N-1]^3)+\\
+q^4(2[N-1]([N-1][N-2]+[2])+3[2][N-1]^3)-q^5[2][N-1]([N-1][N-2]+[2]) \Big )=\\
=[N] \left ( 1-A^2(A^2+1)\{ Aq\}\{A/q\} \right ).
\end{aligned}
\end{equation}

\subsection{The four-vertex virtual twist knot}

\begin{picture}(300,110)(-50,-20)
\put(0,0){\line(1,0){60}}
\qbezier(0,0)(0,80)(30,55)
\qbezier(60,0)(60,80)(30,55)
\qbezier(30,30)(50,20)(20,0)
\qbezier(30,30)(10,20)(40,0)
\qbezier(30,30)(50,45)(30,55)
\qbezier(30,30)(10,45)(30,55)

\qbezier(20,0)(10,-10)(30,-10)
\qbezier(40,0)(50,-10)(30,-10)
\put(20,0){\circle*{3}}
\put(40,0){\circle*{3}}
\put(30,7){\circle*{3}}
\put(30,30){\circle*{3}}
\put(30,55){\circle{5}}

\put(12,0){\vector(1,0){2}}
\put(31,-10){\vector(1,0){2}}
\qbezier(260,0)(230,-10)(200,0)
\qbezier(260,0)(260,80)(235,65)
\qbezier(200,0)(200,80)(225,65)
\qbezier(235,65)(210,47)(230,47)
\qbezier(225,65)(250,47)(230,47)
\put(230,10){\circle{15}}
\put(230,33){\circle{15}}
\put(230,61){\circle{5}}

\put(230,21){\circle*{2}}

\put(230,44){\circle*{2}}
\put(220,2){\circle*{2}}
\put(240,2){\circle*{2}}

\end{picture}

\noindent
For the similar virtual knot the primary and main classical  hypercubes are:

\begin{equation}
\begin{array}{ccccc}
 & &  1 \\
  & 2 & 1 & 2 \\
& 2 & 1 & 2 \\
 3 & & & &1\\
& 2 & 3 & 2\\
& 2 & 1 & 2\\

 & & 1  \\
\end{array}
\longrightarrow
\begin{array}{c|c|c|c|c}

 & & N(N-1)^2  \\
 &  N^2(N-1) & N(N-1)^2 &\underline{ N(N-1)(N-3)}\\
 &  N^2(N-1) & N(N-1)^2 &2N(N-1)^2\\

\boxed{ N^3} & & & & \underline{2N(N-1)(N-3)}\\

 &  N^2(N-1) & 2N^2(N-1) & 2N(N-1)^2\\
 &\boxed{ \boxed{  N^2(N-1) }} & N(N-1)^2 &\underline{ N(N-1)(N-3)}\\
 & & N(N-1)^2  \\
\end{array}
\end{equation}
\noindent We can find out the quantization of the underlined item using the requirement that the coloring for the initial double-boxed initial vertex is the unknot:

\begin{picture}(300,110)(-50,-20)
\put(0,0){\line(1,0){60}}
\qbezier(0,0)(0,80)(30,55)
\qbezier(60,0)(60,80)(30,55)
\qbezier(30,30)(50,20)(20,0)
\qbezier(30,30)(10,20)(40,0)
\qbezier(30,30)(50,45)(30,55)
\qbezier(30,30)(10,45)(30,55)

\qbezier(20,0)(10,-10)(30,-10)
\qbezier(40,0)(50,-10)(30,-10)
\put(20,0){\circle*{3}}
\put(40,0){\circle{3}}
\put(30,6){\circle*{3}}
\put(30,30){\circle*{3}}
\put(30,55){\circle{5}}

\put(12,0){\vector(1,0){2}}
\put(31,-10){\vector(1,0){2}}
\end{picture}

\begin{equation}
\begin{array}{l}
-q^{2N-3}[N] \Big ( [N][N-1]-q([N]^2+2[N-1]^2+[2][N][N-1])+
\\[0.2cm]
+q^2(3[N][N-1]+2[2][N-1]^2+[N-1]([N-2]-1))-
\\[0.2cm]
-q^3(3[N-1]^2+[2][N-1]([N-2]-1))+q^4[N-1]([N-2]-1) \Big )=[N],
\end{array}
\end{equation}
i.e. the relevant quantization rule in this case is $ \underline{(N-1)(N-2)}\rightarrow  \underline{ [N-1]([N-2]-1)} $.
Now we can find HOMFLY for original (all black) coloring:
\begin{equation}
\begin{array}{l}
q^{4(N-1)}[N] \Big ([N]^2-4q[N][N-1]+q^2(5[N-1]^2+[2][N][N-1])-
\\[0.2cm]
-q^3(2[2][N-1]^2+2[N-1]([N-2]-1))+q^4[2][N-1]([N-2]-1) \Big )=
\\[0.2cm]
\hfill=[N]\left (1- (A/q)(A^2+Aq+1)(Aq-1)\{A/q\} \right ).
\end{array}
\end{equation}

\section{On differential expansion for the fundamental HOMFLY   \label{diffvirt}}

\subsection{The claim \label{cladi}}

Let us now return to eq.(\ref{difvirtfund}), which we just checked
in a number of examples.
In \cite{IMMMfe,evo,artdiff} a powerful idea of differential expansion
was introduced for knot polynomials -- and proved to be useful
for constructing {\it colored} polynomials, i.e. those in
non-fundamental (primarily, symmetric and antisymmetric) representations.
The starting point in \cite{IMMMfe} was exactly like (\ref{difvirtfund}):
the statement that the deviation $h_{_\Box}-1$  of reduced (normalized)
HOMFLY polynomial  $h_{_\Box} = [N]^{-1}
H_{_\Box}$ from unity vanishes at $A=q^{\pm 1}$.
This should be true in the fundamental representation,
in higher symmetric representations
\begin{equation}
h_{[r]}-1 \sim \{Aq^r\}\{A/q\}
\label{diffhi}
\end{equation}
and the coefficient in this proportionality relation
is further decomposed in a similar way
\cite{IMMMfe}-\cite{diffappsfirst}.

Like $h_{_\Box}-1$ for ordinary knots, the difference $v_{_\Box}-1$
also vanishes at $A=q^{\pm 1}$,
however, as clear from (\ref{difvirtfund}), for virtual knots this happens
in a somewhat asymmetric symmetric way: while for the
ordinary knots
\begin{equation}
h_{_\Box}-1 \sim \{Aq\}\{A/q\} \sim (q^{N+1}-q^{-1-N})(q^{N-1}-q^{1-N})\sim(q^{N+1}-1)(q^{N-1}-1),
\end{equation}
for virtual ones
\begin{equation}
\boxed{v_{\Box}-1  \sim (Aq-1)\{A/q\} } \sim \{\sqrt{Aq}\}\{A/q\}
\sim (q^{N+1}-1)(q^{N-1}-q^{1-N}) \sim(q^{N+1}-1)(q^{N-1}-1).
\end{equation}

In this section we provide certain evidence that this happens not only
for the series of virtual twist knots, but in more general situation,
presumably -- {\it always}.
Moreover, this property is true not only for reduced HOMFLY for knots,
it seems also to hold for {\it unreduced} HOMFLY for links.

\subsection{Small virtual knots}

Fundamental HOMFLY polynomials, found in \cite{MMMvirt} for the first few virtual knots
from the table \cite{virtable}, can be rewritten as:
\begin{equation}
\begin{array}{l}
V^{2.1}_{_\Box} = -\frac{A^3}{q}+\frac{A^2}{q^2}+Aq
= 1 -\frac{A}{q} \, \boxed{(Aq-1)\{A/q\}}\,,
\\[0.2cm]
V^{3.1}_{_\Box} = 1,
\\[0.2cm]
V^{3.2}_{_\Box} = A^2-\frac{A}{q}-q^2+1+\frac{q}{A} = 1 +\boxed{(Aq-1)\{A/q\}}\,,
\\[0.2cm]
V^{3.3}_{_\Box} = V^{2.1}_{_\Box},
\\[0.2cm]
V^{3.4}_{_\Box} = V^{3.2}_{_\Box},
\\[0.2cm]
V^{3.5}_{_\Box} =  -A^4+(q^2+q^{-2})A^2 = 1-A^2\,\boxed{\boxed{\{Aq\}\{A/q\}}}= 1 - \frac{A}{q}(Aq+1)\,\boxed{(Aq-1)\{A/q\}} ,
\\[0.2cm]
V^{3.6}_{_\Box} = V^{3.5}_{_\Box}=H^{[2,3]}_{_\Box},
\\[0.2cm]
V^{3.7}_{_\Box} = 1,
\\
\ldots
\end{array}
\end{equation}
All these knots are topologically distinct, however there are many
coincidences between their fundamental HOMFLY polynomials.

In application to the fundamental HOMFLY the differential hierarchy structure
is the appearance of the {\it universal} factor in a box -- presumably,
for arbitrary virtual knot.
For ordinary knots this property is enhanced:
the universal factor is promoted to that in the double box --
in our small table this is represented by the example of $3.6$,
which is actually an ordinary trefoil $3_1$, i.e. the ordinary
2-strand torus knot $[2,3]$.

\subsection{2-strand torus knots}

From (\ref{2stratorus}) we get:
\begin{equation}
\begin{array}{l}
h_{_\Box}^{[2,3]} = 1 - q^{2N}\,\boxed{\boxed{\{Aq\}\{A/q\}}},
\\[0.3cm]
h_{_\Box}^{[2,5]} = 1 - q^{2N}\,\boxed{\boxed{\{Aq\}\{A/q\}}}\Big(1+(q^2+q^{-2})q^{2N}\Big),
\\[0.3cm]
h_{_\Box}^{[2,7]} = 1 - q^{2N}\,\boxed{\boxed{\{Aq\}\{A/q\}}}\Big(1+(q^2+q^{-2})q^{2N}
+(q^4+1+q^{-4})q^{4N}\Big),
\\[0.3cm]
h_{_\Box}^{[2,9]} = 1 - q^{2N}\,\boxed{\boxed{\{Aq\}\{A/q\}}}\Big(1+(q^2+q^{-2})q^{2N}
+(q^4+1+q^{-4})q^{4N}+(q^6+q^2+q^{-2}+q^{-6})q^{4N}\Big),
\\
\ldots
\end{array}
\end{equation}
and from (\ref{virt2stratorus}) ---
\begin{equation}
\begin{array}{l}
v_{_\Box}^{[2,3]} = V_{_\Box}^{2.1} = 1 -  q^{N-1}\, \boxed{(q^{N+1}-1)\{A/q\}},
\\[0.2cm]
v_{_\Box}^{[2,5]} = 1 -  q^{N-1}\, \boxed{(q^{N+1}-1)\{A/q\}}\Big((1+q^{N+1})+(q^2+q^{-2})q^{2N}\Big),
\\[0.2cm]
v_{_\Box}^{[2,7]} = 1 - q^{N-1}\, \boxed{(q^{N+1}-1)\{A/q\}}
\left\{(1+q^{N+1})\Big(1+(q^2+q^{-2})q^{2N}\Big)
+(q^4+1+q^{-4})q^{4N}\right\},
\\[0.2cm]
v_{_\Box}^{[2,9]} = 1 -  q^{N-1}\, \boxed{(q^{N+1}-1)\{A/q\}}
\Big\{(1+q^{N+1})\Big(1+(q^2+q^{-2})q^{2N}
+(q^4+1+q^{-4})q^{4N}\Big)+
\\[0.2cm] \hfill
+(q^6+q^2+q^{-2}+q^{-6})q^{4N} \Big\},
\\
\ldots
\end{array}
\end{equation}
This extends differential hierarchy from twist to the simplest torus family.
However, the coefficients in front of the universal factors in the torus case
are considerably more complicated than for twist knots -- this gets even
worse in the colored case \cite{artdiff}.

\subsection{Parallel 2-strand links}

Moreover, the structure survives also for two-component torus links,
only now we take unreduced HOMFLY.
From the same (\ref{2stratorus}) and (\ref{virt2stratorus}) we deduce
\begin{equation}
H_{_\Box}^{[2,n]}=q^{Nn} \left ( q^{-n} \left ( [N]^2-\frac{[N][N-1]}{[2]} \right )+q^n \frac{[N][N-1]}{[2]} \right )
\end{equation}
and
\begin{equation}
V_{_\Box}^{[2,n+1]} =q^{Nn} \left ( q^{-n} \left ( [N]+\frac{[N][N-1]}{[2]} \right )+q^n \frac{[N][N-1]}{[2]} \right ).
\end{equation}
These give the following answers for simplest 2-strand links:
\begin{equation}
\begin{array}{l}
H_{_\Box}^{[2,2]} = 1+q^{2N}\,\boxed{\boxed{\{Aq\}\{A/q\}}} { q^2}/{(q^2-1)^2}, \\[0.3cm]
H_{_\Box}^{[2,4]} = 1+ q^{2N}\,\boxed{\boxed{\{Aq\}\{A/q\}}} {(q^{2N}(q^4-q^{2}+q)-q^4+2q^2-1)}/{(q^2-1)^2}, \\[0.3cm]
H_{_\Box}^{[2,6]} = 1+ q^{2N}\,\boxed{\boxed{\{Aq\}\{A/q\}}} (q^{4N}(q^8-q^6+q^4-q^2+1)+q^{2N}(-q^8+2q^6-2q^4+2q^2-1)-
\\[0.3cm] \hfill -q^6+2q^4-q^2)/{q^2(q^2-1)^2}, \\
\ldots
\end{array}
\end{equation}

\noindent and

\begin{equation}
\begin{array}{l}
V_{_\Box}^{[2,2]} =  1 + q^{N-1}\, \boxed{(Aq-1)\{A/q\}} ({q^2+q^{3-N}})/{(q^2-1)^2}, \\[0.2cm]
V_{_\Box}^{[2,4]} =  1 + q^{N-1}\, \boxed{(Aq-1)\{A/q\}} ({q^{2N}(q^4-q^{2}+q)+q^{N+3}-q^4+2q^2-1})/{(q^2-1)^2}, \\[0.2cm]
V_{_\Box}^{[2,6]} =  1 + q^{N-1}\, \boxed{(Aq-1)\{A/q\}} \Big(-q^6+2q^4-q^2+q^N(-q^7+2q^5-q^3)+
\\[0.2cm]
+q^{2N}(-q^8+2q^6-2q^4+2q^2-1)+q^{3N}(q^7-q^5+q^3)+q^{4N}(q^8-q^6+q^4-q^2+1)\Big)/q^2(q^2-1)^2,\\
\ldots
\end{array}
\end{equation}

\subsection{Antiparallel 2-strand links}

Similarly, for antiparallel links we obtain from (\ref{antiparrlink}) and (\ref{antiparlink}),
again for unreduced HOMFLY:
\begin{equation}
H^{2,2k}_{_{\Box\times\overline{\Box}}} =
1 + A^{2k}[N+1][N-1] =
1 + \frac{A^{2k}\,\boxed{\boxed{\{Aq\}\{A/q\}}}}{(q-q^{-1})^2}
\end{equation}
and
\begin{equation}
V^{2,2k}_{_{\Box\times\overline{\Box}}} =
1 + A^{2k-1}\Big([N]+1\Big)[N-1] =
1 + \frac{A^{2k-2}\,\boxed{(Aq-1)\{A/q\}}}{(q-q^{-1})^2}\cdot  (q^{N-1}+1).
\end{equation}

It deserves noting, that at $q=1$   un-reduced HOMFLY for $l$-component
links are equal to $N^l$.
In this limit also $A=1$, but the ratio
$\frac{\{Aq^{\pm 1}\}}{\{q\}}\longrightarrow N\pm 1$,
thus both terms in the differential decomposition contribute in this limit.

\subsection{Differential expansion for Jones polynomials}

Above examples exhaust all the virtual knots with already known HOMFLY polynomials.
For a wider class of examples the Jones polynomials (i.e. HOMFLY at $N=2$)
can be found in \cite{virtable}.

For fundamental reduced Jones our differential hierarchy structure implies
that the difference $J_{_\Box}-1$ is divisible by $(Aq-1)\{A/q\} \longrightarrow
(q^3-1)(q^2-1)$ -- but not by $(q^6-1)(q^2-1)$ as it does for the ordinary knots.
This is indeed the case for all the examples discussed in \cite{MMMvirt} and for several checked examples from \cite{virtable}, The coefficient $g_{_\Box}^{\cal L}$ introduced using
\begin{equation}
J^{\cal L}_{_\Box} = 1 + g_{_\Box}^{\cal L}(q)\,\boxed{(q^3-1)\left(q-q^{-1}\right)}
\end{equation}
is equal (after a substitution $q^{-1/2}\rightarrow q$ in \cite{virtable}) to:
\begin{equation}
\begin{array}{l}
g_{_\Box}^{4.1} = q^6-q^4-q^3-q, \\[0.2cm]
g_{_\Box}^{4.2} = -q^{-1}-q^{-2}= -q^{-2}(q+1), \\[0.2cm]
g_{_\Box}^{4.3} = g_{_\Box}^{4.1},  \\[0.2cm]
g_{_\Box}^{4.4} = - q, \\
\ldots \\
g_{_\Box}^{4.49} = g_{_\Box}^{4.4},  \\[0.2cm]
g_{_\Box}^{4.50} = -q^2-q = -q(q+1),  \\[0.2cm]
\ldots \\
g_{_\Box}^{4.107} = 0, \\[0.2cm]
g_{_\Box}^{4.108} = \boxed{\boxed{1+q^{-3}}}=q^{-3}(q^3+1).
\ \ \ \text{This is the ordinary knot}\ 4_1. \\
\end{array}
\end{equation}

\section{On colored HOMFLY for virtual knots
\label{colored} }

\subsection{Conjectures}

Since the very beginning in \cite{IMMMfe} differential expansion
was used to promote fundamental HOMFLY in two directions:
to higher representations (at least, (anti)symmetric) and to superpolynomials.
It is a very interesting question, whether this can be also done
for virtual knots and links.
In this section we say just a few words about the first of these options.

As already mentioned in section \ref{cladi},
differential expansion for colored polynomials forms possesses a deep
hierarchical structure \cite{artdiff}, but it starts from (\ref{diffhi}) --
a direct analogue of what we discussed in the previous section
for the fundamental representation.
However, even this first step can be a problem when we switch to virtual knots.
Relation (\ref{diffhi}),
\begin{equation}
h_{[r]}-1 \sim \{Aq^r\}\{A/q\},
\label{diffhi2}
\end{equation}
could be hypothetically promoted to
\begin{equation}
v_{[r]}-1 \ \stackrel{?}{\sim}\ (Aq^r-1)\{A/q\},
\label{diffvi}
\end{equation}
but the problem is to define what $v_{[r]}$ could be --
because representation-theory structure is broken by the
sterile vertices.
The question is -- to what extent it is broken:
whether just {\it internal} structure is deformed
{\it within} irreducible representations or
different representations are intermixed.
This section describes simple observations, which can help
thinking about this issue.

While definition of {\it colored} HOMFLY $V_R$
with arbitrary Young diagram $R$ is not at all straightforward
for virtual knots and links, one can easily define
the {\it cabled} HOMFLY $V_{\Box^{\otimes r}}$ --
by substituting a line in the knot by an $r$-strand cable:

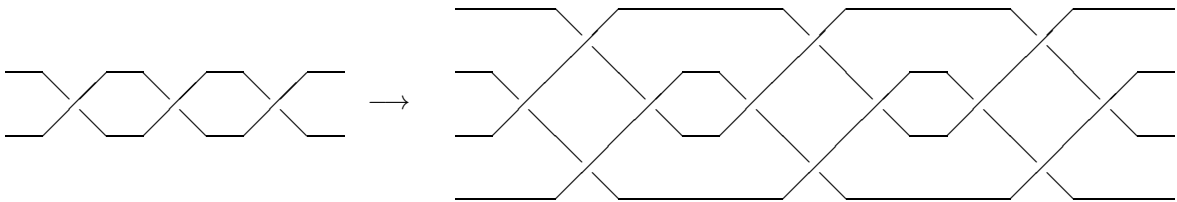
\begin{picture}(360,100)(0,-37)
\put(0,0){\line(1,0){14}}
\put(0,24){\line(1,0){14}}
\put(14,0){\line(1,1){24}}
\put(14,24){\line(1,-1){10}}
\put(28,10){\line(1,-1){10}}
\put(38,0){\line(1,0){14}}
\put(38,24){\line(1,0){14}}
\put(52,0){\line(1,1){24}}
\put(52,24){\line(1,-1){10}}
\put(66,10){\line(1,-1){10}}
\put(76,0){\line(1,0){14}}
\put(76,24){\line(1,0){14}}
\put(90,0){\line(1,1){24}}
\put(90,24){\line(1,-1){10}}
\put(104,10){\line(1,-1){10}}
\put(114,0){\line(1,0){14}}
\put(114,24){\line(1,0){14}}
\put(145,12){\makebox(0,0)[cc]{$\longrightarrow$}}
\put(170,-24){\line(1,0){38}}
\put(170,0){\line(1,0){14}}
\put(170,24){\line(1,0){14}}
\put(170,48){\line(1,0){38}}
\put(208,-24){\line(1,1){48}}
\put(184,0){\line(1,1){48}}
\put(184,24){\line(1,-1){10}}
\put(198,10){\line(1,-1){20}}
\put(222,-14){\line(1,-1){10}}
\put(208,48){\line(1,-1){10}}
\put(222,34){\line(1,-1){20}}
\put(246,10){\line(1,-1){10}}
\put(232,-24){\line(1,0){62}}
\put(256,0){\line(1,0){14}}
\put(256,24){\line(1,0){14}}
\put(232,48){\line(1,0){62}}
\put(270,0){\line(1,1){48}}
\put(294,-24){\line(1,1){48}}
\put(270,24){\line(1,-1){10}}
\put(284,10){\line(1,-1){20}}
\put(308,-14){\line(1,-1){10}}
\put(294,48){\line(1,-1){10}}
\put(308,34){\line(1,-1){20}}
\put(332,10){\line(1,-1){10}}
\put(318,-24){\line(1,0){62}}
\put(342,0){\line(1,0){14}}
\put(342,24){\line(1,0){14}}
\put(318,48){\line(1,0){62}}
\put(356,0){\line(1,1){48}}
\put(380,-24){\line(1,1){48}}
\put(356,24){\line(1,-1){10}}
\put(370,10){\line(1,-1){20}}
\put(394,-14){\line(1,-1){10}}
\put(380,48){\line(1,-1){10}}
\put(394,34){\line(1,-1){20}}
\put(418,10){\line(1,-1){10}}
\put(404,-24){\line(1,0){38}}
\put(428,0){\line(1,0){14}}
\put(428,24){\line(1,0){14}}
\put(404,48){\line(1,0){38}}
\end{picture}

\noindent
and taking HOMFLY of this virtual link.
It is sometime suggested to consider $\widetilde {V}_{\Box^{\otimes r}}$
-- a HOMFLY for polynomial for a link with additional twistings
inside the cable, added to make the linking numbers of every (?) two strands
in the cable vanishing.
Similar twistings are used to make projections on particular representations
$R$ in the decomposition
\be
H_{_{\Box^{\otimes r}}} = \sum_{R,\ |R|=r} H_R
\ee
of cabled into colored HOMFLY for ordinary knots and links.
Immediate question is if a similar decomposition can exist
in the virtual case:
\be
V_{_{\Box^{\otimes r}}} \ \stackrel{?}{=}\ \sum_{R,\ |R|=r} V_R.
\label{Vdeco}
\ee

A possible obstacle is that the sterile vertex can provide
non-vanishing transitions between different representations:

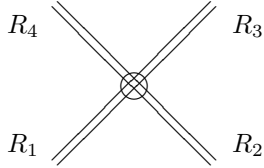
\begin{picture}(300,100)(-150,-50)
\put(-31,-28){\line(1,1){60}}
\put(-29,-30){\line(1,1){60}}
\put(31,-28){\line(-1,1){60}}
\put(29,-30){\line(-1,1){60}}
\put(0,-0){\circle{10}}
\put(-48,-25){\mbox{$R_1$}}
\put(-48,20){\mbox{$R_4$}}
\put(37,20){\mbox{$R_3$}}
\put(37,-25){\mbox{$R_2$}}
\end{picture}

\noindent
even if there is a single virtual vertex in the knot, so that
$R_3=R_1$ and $R_4=R_2$ because of topology of the rest of the graph,
$R_2$ can still be different from $R_1$ and in that case (\ref{Vdeco})
should then be substituted by a far more complicated decomposition
\be
V_{_{\Box^{\otimes r}}} \ \stackrel{?}{=}\ \sum_{R_1,R_2\ |R_i|=r} V_{R_1R_2}
\label{Vdeco2}
\ee
and even worse -- when the number of sterile vertices increases

To make the choice between (\ref{Vdeco}) and (\ref{Vdeco2}) phenomenologically
one needs to know examples of cabled HOMFLY.
In the very simplest case of $2.1$ and $r=2$ the relevant link diagram
has eight vertices -- and still needs to be calculated
(after all, the present paper is just the second after \cite{MMMvirt},
where HOMFLY for virtual knots are considered).
Known, however, are many cabled Jones \cite{virtable}.

Since Jones polynomials in these tables are reduced, while (\ref{Vdeco})
involves unreduced ones, the relation should be rewritten as follows:
\be
(q+q^{-1})J_{_{\Box\times\Box}} \ \stackrel{?}{=}\ (q^2+1+q^{-2})J_{[2]} + 1,
\label{Jdeco2}
\ee
\be
(q+q^{-1})J_{_{\Box\times\Box\times\Box}} \ \stackrel{?}{=}\
(q^3+q+q^{-1}+q^{-3})J_{[3]} + 2\,(q+q^{-1})J_{_\Box},
\label{Jdeco3}
\ee
\be
\ldots
\nn
\ee
(we remind that the substitution $\ \ q^{-1/2}\longrightarrow q\ \ $ should be made
in \cite{virtable}, and also the signs of even cabled polynomials needs be reversed).
In these formulae we used the fact that for $N=2$ all representations
are symmetric, in particular $[11]\equiv [0]$ and $[21]\equiv [1]=\Box$,
while $[111]$ does not contribute at all -- and dimensions of representations
are just the quantum numbers $D_r(N=2) = [r+1]$.
The extra multiplicity $2$ in (\ref{Jdeco3}) comes from decomposition
rule $\ [1]\times[1]\times[1] = [3] + 2\cdot [21] + [111]$.

If (\ref{Jdeco2}) and (\ref{Jdeco3}) were true, one could use them to
define $J_{[2]}$ and $J_{[3]}$ -- and then, iteratively, next $J_{[r]}$ --
from known cabled Jones.
The minimal criterium of truth is that the resulting quantities are indeed
Laurent polynomials in $q$ -- i.e. the difference
$(q+q^{-1})J_{_{\Box\times\Box}}-1$ is divisible by $(q^2+1+q^{-2})=[3]$ and so on.

\subsection{Check for ordinary knots}

This is of course true for ordinary knots, i.e. for $3.6$ and $4.108$ of
\cite{virtable}, which are respectively the ordinary trefoil and
figure eight knots:
\begin{equation}
\begin{array}{l}
J_{[2]}^{3.6} = \cfrac{[2]J_{_{\Box\times\Box}}^{3.6}-1}{[3]} = q^{22}-q^{20}-q^{18}+q^{16}-q^{14}+q^{10}+q^4,
\\ \\
J_{[3]}^{3.6} = -q^{42}+q^{40}+q^{38}-q^{34}+q^{30}-q^{28}-q^{26}+q^{22}-q^{20}+q^{14}+q^6
\end{array}
\end{equation}
and
\begin{equation}
\begin{array}{l}
J_{[2]}^{4.108} = \frac{[2]J_{_{\Box\times\Box}}^{4.108}-1}{[3]} = q^{12}-q^{10}-q^8+2q^6-q^4-q^2+3-q^{-2}-q^{-4}+2q^{-6}-q^{-8}-q^{-10}+q^{-12},
\\ \\
J_{[3]}^{4.108} =q^{24}-q^{22}-q^{20}+2q^{16}-2q^{12}+3q^8-3q^4+3-3q^{-4}+3q^{-8}-2q^{-12}+2q^{-16}-q^{-20}-q^{-22}+q^{-24}.
\end{array}
\end{equation}
These are the well known correct expressions, moreover
\begin{equation}
\begin{array}{l}
J_{[2]}^{3.6} - 1 \sim (q^4-1)(q^2-1), \\
J_{[2]}^{4.108} - 1 \sim (q^4-1)(q^2-1), \\
\\
J_{[3]}^{3.6} - 1 \sim (q^5-1)(q^2-1), \\
J_{[3]}^{4.108} - 1 \sim (q^5-1)(q^2-1), \\
\ldots
\end{array}
\end{equation}
as conjectured in (\ref{diffvi}).
However, in these two cases this is just a corollary of
(\ref{diffhi2}).

\subsection{Validation of (\ref{Jdeco2}) and support to (\ref{diffvi}) for $r=2$}

A real surprise is that (\ref{Jdeco2}) works not only for ordinary knots but also for {\it virtual} ones from \cite{virtable}.
Moreover, so defined $J_{[2]}$ do indeed possess the property (\ref{diffvi}),
$J_{[2]}-1\sim (q^4-1)(q^2-1)$. Some examples are listed int the table \ref{tj}. The coefficients are defined as
\begin{equation}
\begin{array}{c}
J_{_\Box}-1 = g_{_\Box}\cdot(q^3-1)(q-q^{-1}),
\\[0.4cm]
J_{[2]} - 1 = g_{[2]}\cdot (q^4-1)(q-q^{-1}) =
\Big([2]\,g_{_\Box}+g^{(2)}_{[2]}\cdot(q^4-1)(q-q^{-1}) \Big),
\end{array}
\end{equation}
while colored Jones  $J_{[2]}$ is extracted from
\begin{equation}
(q+q^{-1})J_{_{\Box\times\Box}}  + (q^2+1+q^{-2}) J_{[2]} + 1 = 0.
\end{equation}

\subsection{Violation of (\ref{Jdeco3})}

However, the same trick does {\it not} work for (\ref{Jdeco3}). One cannot extract $J_{[3]}$ for virtual knots from the cabled polynomials from \cite{virtable}. For example the conjecture (\ref{Jdeco3}) gives for the knot $2.1$
\begin{equation}
\begin{array}{r}
J_{[3]}^{2.1}\stackrel{?}{=}-\cfrac{1}{q^2+q^{-2}}(q^{29}-q^{27}+q^{25}-q^{23}-3q^{22}-2q^{21}-q^{20}+q^{19}+3q^{18}+
\\[0.4cm]
+2q^{17}+4q^{16}+4q^{15}-2q^{13}-3q^{12}-q^{11}-q^{10}-2q^9-q^8+q^6-q^4).
\end{array}
\end{equation}
Likewise for $3.1$
\begin{equation}\!\!\!\!\!\!
\begin{array}{r}
J_{[3]}^{3.1}\stackrel{?}{=}-\cfrac{1}{q^2+q^{-2}}(q^{21}-q^{19}-2q^{18}-5q^{17}-7q^{16}-14q^{15}-14q^{14}-10q^{13}-7q^{12}+2q^{11}+13q^{10}+26q^{9}+
\\[0.4cm]
+33q^{8}+24q^{7}+19q^{6}+7q^{5}-7q^{4}-11q^{3}-21q^{2}
-13q-12-4q^{-1}+q^{-2}-2q^{-3}+q^{-4}+2q^{-6}-q^{-8}).
\end{array}
\end{equation}
As one can see these answers are not polynomials and thus can hardly be interpreted as the colored Jones polynomials.

It is an intriguing question why (\ref{Jdeco2}) works and (\ref{Jdeco3})
does not -- as we explained, it would be natural to expect that neither one does.
Perhaps, the success with the first symmetric representation $[2]$
will be limited to Jones ($N=2$) case, when antisymmetric representation $[11]$
trivializes to a singlet.
Perhaps, the sterile vertex fails to mix a singlet with $[2]$,
but does so with $[11]$ and $[2]$ in general.

Note also that we did not really distinguish $\tilde J$ from $J$ (i.e. Jones polynomials for cables with additional crossings corresponding to the vanishing linking numbers and Jones polynomial for plain cable),
while, strictly speaking (\ref{Jdeco2}) and (\ref{Jdeco3}) are about
$J$ (for $\tilde J$ extra powers of $q$ can appear for twisting insertions),
while the data in \cite{virtable} is presumably for $\tilde J$. It is unclear whether thiis kind of modification could cure (7.7) and provide a reasonable ``definition'' of $J_{[3]}$.

\newpage
\begin{landscape}
\begin{table}
$$
\begin{array}{|c||c|c||c|c|c|}
\hline
\\
\text{knot} & J_{_\Box} &  g_{_\Box}=\frac{J_{_\Box}-1}{(q^3-1)(q-q^{-1})} &
J_{_{\Box\times\Box}}& J_{[2]}\ \text{from}\ (\ref{Jdeco2}) &
g_{[2]} = \frac{J_{[2]}-1}{(q^4-1)(q-q^{-1})} \\
&&&&&\\
\hline
&&&&&\\
2.1 &-q^5+q^3+q&-q& q^{15}-q^{13}-q^{11}-q^9+ & q^{14}-q^{12}-2q^{10}+ &  q^9-q^5-q^3-q \\
&&& +q^7+2q^5+q    &+ q^8+q^6+q^4 & \\
&&&&&\\
\hline
&&&&&\\
3.1 &1&0&-q^9-q^7+q^5+2q^3+2q-q^{-3}& -q^8-q^6+2q^4+2q^2-q^{-2} & -q^3-2q-q^{-1} \\
&&&&&\\
\hline
&&&&&\\
3.2 &q^4-q^2-q+1+q^{-1}&1& q^{13}-q^{11}-q^9+2q^5+&q^{12}-q^{10}-2q^8+2q^6+& q^7-q^3+q^{-1}+q^{-3}\\
&&                       &+q^3-q+q^{-5}           & +2q^4-q^2-1+q^{-4}&\\
&&&&&\\
\hline
&&&&&\\
3.3 &-q^5+q^3+q^2-q&-q&-q^{17}+q^{11}+q^5+q&-q^{16}+q^{12}+q^4&-q^{11}-q^9-q^7-q^5-q^3-q\\
&&&&&\\
\hline
&&&&&\\
3.4 &q^4-q^2-q+1+q^{-1}&1&q^{13}-2q^{11}-2q^9+q^7+4q^5+&q^{12}-2q^{10}-3q^8+4q^6+&q^7-q^5-3q^3-q+q^{-1}+q^{-3}\\
&&                       &+2q^3-2q-q^{-1}+q^{-5}       & +4q^4-2q^2-2+q^{-4}&\\
&&&&&\\
\hline
&&&&&\\
3.5 &-q^8+q^6+q^2&-q^4-q&q^{23}-q^{21}+q^{19}-q^{17}-2q^{15}&q^{22}-q^{20}+q^{16}-4q^{14}+&q^{17}+q^{13}+q^{11}-2q^9-\\
&&                      &-q^{13}+q^{11}+3q^9+q              &+4q^{10}-q^6+q^4             &-2q^7-q^5-q^3-q\\
&&&&&\\
\hline
&&&&&\\
3.6 &-q^{8}+q^6+q^2&-q^4-q&q^{23}-q^{21}-q^{17}+q^9+q^5+q&q^{22}-q^{20}-q^{18}+q^{16}-&q^{17}-q^9-q^7-q^5-q^3-q\\
&&                        &                              &-q^{14}+q^{10}+q^4             & \\
&&&&&\\
\hline
&&&&&\\
3.7 &1&0&q^{11}-2q^7-q^5+q^3+&q^{10}-3q^6+3q^2+1-q^{-2}&q^5+q^3-q-q^{-1}\\
&&      &+3q+q^{-1}-q^{-3}   & & \\
&&&&&\\
\hline
\end{array}
$$
\caption{\label{tj} In this table the Jones, cabled Jones and colored Jones in representation [2] are listed. Also provided are coefficients $g_{\Box}$ and $g_{[2]}$.}
\end{table}
\end{landscape}
\newpage


Still another problem is that the deeper levels of differential hierarchy
involve factorization of differences like $(G_{[r]}-[r]G_{_\Box})$
in $h_{[r]}-1 = G_{[r]}\{Aq^r\}\{A/q\}$ -- but
such differences are considerably changed after switching from coefficients
in front of $\{Aq^r\}\{A/q\}$ to those in front of $\{\sqrt{Aq^r}\}\{A/q\}$,
which seem to be relevant in virtual case.

In any case, to handle all these problems there is a big need to calculate
cabled HOMFLY -- Jones is clearly insufficient.
This is a straightforward task in the approach of \cite{DM3}, \cite{MMMvirt}
and the present paper -- the problem is just technical:
to effectively computerize calculations for arbitrary $N$,
as it was once done for Jones \cite{BNcomp}.
It is only needed to include new resolutions, step-by-step construction
of the full quantum hypercube -- on one hand,
and to include link diagrams with sterile vertices -- on another.

\section{Conclusion}

In this paper we provided a new significant evidence that
the formalism of \cite{DM3} allows to define
topologically invariant HOMFLY polynomials
for virtual knots and links -- despite the usual
RT approach is not directly applicable.
We extended the 2-strand torus example from original
paper \cite{MMMvirt} to the families of antiparallel 2-strand braids
and twist knots.
This is important not only for the check of topological invariance
-- there are non-trivial equivalencies within the twist family --
but also they provide new evidence for partial
survival of group-theory structures and their implication,
like the differential-hierarchy structure (which is modified by a
straightforward, still mysterious substitution $\{Aq\}\{A/q\}
\longrightarrow \{\sqrt{Aq}\}\{A/q\}$).
Moreover, these observations open a way to the introduction
of colored (not just cabled) HOMFLY polynomials:
we provided some optimistic evidence about representation $[2]$,
and described immediate problems in the case of representation $[3]$.

Our conclusion is that at least the fundamental HOMFLY polynomials
clearly exist for virtual knots, and perhaps even the RT
method can be modified to describe them and introduce colored HOMFLY.
This however, requires further investigation.
Another open way is to further develop the method of \cite{DM3} --
which now proved capable to provide essentially new results.
From the perspective of virtual knots, an interesting question
is if Khovanov-Rozansky and super- polynomials can also be constructed
for them.

\section*{Acknowledgements}

We are indebted to A.Mironov for attracting our attention to the
subject of virtual knots and to D.Bar-Natan for valuable comments
on the subject. We are also grateful to A.Anokhina, S.Mironov, A.Sleptsov and Ye.Zenkevich for useful discussions.
Our work is partly supported by grant NSh-1500.2014.2 (L.B., A.M. and And.M.), by RFBR grants 13-02-00478 (A.M.),
14-01-00547 (And.M.), 14-01-31395\_young\_a (And.M.), 12-01-00482 (Ant.M.), 14-02-31446\_young\_a (Ant.M.), by joint grants 13-02-91371-ST and 14-01-92691-Ind (A.M. and And.M.), by the Brazil National Counsel of Scientific and Technological Development (A.M.), by the Laboratory of Quantum Topology of Chelyabinsk State University (Russian Federation government grant 14.Z50.31.0020) (And.M.) and by the Dynasty Foundation (And.M.).

\end{document}